\documentclass[10pt,aps,preprint,floatfix,twocolumn]{revtex4}

\usepackage{amsmath, amsthm, amssymb}
\usepackage{graphicx}
\usepackage{subfigure}
\usepackage{ulem}
\numberwithin{equation}{section}

\begin{document}

\title{{Influence of gas compression on flame acceleration in the early
stage of burning in tubes}}

\author{ Damir M. Valiev$^{{\rm a}}$ }
\author{V'yacheslav Akkerman$^{{\rm b}}$}
\author{Mikhail Kuznetsov$^{{\rm c}}$}
\author{Lars-Erik Eriksson$^{{\rm d}}$}
\author{Chung K. Law$^{{\rm a,e}}$}
\author{Vitaly Bychkov$^{{\rm f}}$}

\affiliation{ $^{{\rm a}}$Department of Mechanical and Aerospace Engineering, Princeton University, Princeton, NJ 08544-5263, USA \\
$^{{\rm b}}$Department of Mechanical and Aerospace Engineering, West Virginia University, Morgantown, WV 26506-6106, USA \\
 $^{{\rm c}}$Institute for Nuclear and Energy Technologies, Karlsruhe Institute of Technology, Kaiserstrasse 12, 76131 Karlsruhe, Germany \\
$^{{\rm d}}$Department of Applied Mechanics, Chalmers University of Technology, 41296 G\"{o}teborg, Sweden \\
$^{{\rm e}}$Center for Combustion Energy, Tsinghua University, Beijing 100084, China \\
$^{{\rm f}}$Department of Physics, Ume{\aa} University, 90187 Ume{\aa}, Sweden }

\begin{abstract}
The mechanism of finger flame acceleration at the early stage of
burning in tubes was studied experimentally by Clanet and
Searby~[Combust. Flame 105: 225 (1996)] for slow propane-air
flames, and elucidated analytically and computationally by Bychkov et al.~[Combust.
Flame 150: 263 (2007)] in the limit of incompressible flow.
We have now analytically,
experimentally and computationally studied the finger flame acceleration for fast burning flames, when the gas
compressibility assumes an important role. Specifically, we have first developed a  theory
through small Mach number expansion up to the first-order terms,
 demonstrating that gas compression reduces the acceleration
rate and the maximum flame tip velocity, and thereby moderates the finger flame acceleration
noticeably. This is an important quantitative correction to previous theoretical analysis. We have also conducted experiments for hydrogen-oxygen mixtures
with considerable initial values of the Mach number, showing finger flame acceleration with the
acceleration rate much smaller than those obtained previously for
hydrocarbon flames. Furthermore, we have performed numerical simulations for a
wide range of initial laminar flame velocities, with the results
substantiating the experiments. It is shown that  the theory is in good
quantitative agreement with numerical simulations for small gas compression (small initial flame velocities). Similar to
previous works, the numerical simulation shows that finger
flame acceleration is followed by the formation of the ``tulip'' flame, which indicates termination of the early acceleration process.

\textbf{Keywords}: premixed flames, flame acceleration, compressibility, finger flames, tulip flames

\end{abstract}

\maketitle
 \thispagestyle{empty}


\section*{Nomenclature}
\vskip-0.3cm
\begin{tabular}{p{2cm} l}
$c_S$ & sound speed \\
$C_P$ & heat capacity at constant pressure\\
$C_V$ & heat capacity at constant volume \\
$E_a$ & activation energy \\
$Le $ & Lewis number\\
$L_f$ & flame thickness\\
$L_w$ & flame length in a 2D configuration \\
$Ma$ & initial flame propagation Mach number \\
$P$ & pressure\\
$Pr$ & Prandtl number\\
$q_i$ & energy diffusion vector\\
$Q$ & energy release from the reaction\\
$R$ & tube radius (channel half-width) \\
$\overline{R} $ & universal gas constant\\
$Sc $ & Schmidt number\\
$S_L$ & unstretched laminar burning velocity \\
$t$ & time \\
$T $ & temperature\\
$U, u$ & velocity \\
$v $ & dimensionless axial velocity\\
$w $ & dimensionless radial velocity\\
$x,\ r $ & radial coordinate \\
$Y $ & mass fraction of fuel\\
$Z,\ z$ & axial coordinate \\
$\alpha$ & auxiliary constant\\
$\gamma $ & adiabatic index \\
$\varepsilon $ & total energy per unit volume\\
$\zeta_{i,j} $ & stress tensor\\
$\eta $ & dimensionless radial coordinate\\
\end{tabular} 
\begin{tabular}{p{2cm} l}
$ $ &  \\
$\Theta $ & gas expansion ratio  \\
$\vartheta $ & instantaneous expansion factor \\
$\mu $ & dynamical viscosity\\
$\xi $ & dimensionless axial coordinate\\
$\rho$ & density \\
$\sigma$ & scaled acceleration rate \\
$\tau $ & scaled time ($\tau = S_{L} t / R$) \\
$\tau_R $ & factor of time dimension \\
$ $ &  \\
\end{tabular}

\hskip-0.3cm
\textit{Subscripts and other designations} 
\vskip0.2cm
\hskip-0.3cm\begin{tabular}{p{2cm} l}
$0 $ & initial\\
$pl $ & planar geometry\\
$axi $ & axisymmetric geometry\\
$b$ & burnt gas\\
$C$ & modified\\
$f $ & flame, flame skirt\\
$s $ & sidewall\\
$s, f $ & just ahead of the flame skirt \\
$s, b, f$ & in the burnt gas at the flame skirt\\
$sph$ & instant/locus of shape change  \\
$tip$ & flame tip \\
$u$ & unburnt mixture\\
$w $ & wave\\
$wall$ & flame touches side walls of tube \\
$\tilde{ }$ & scaled value \\
\end{tabular}

\newpage

\section{Introduction}
\label{Sec.intro}

The spontaneous acceleration of a flamefront propagating from the
closed end of a tube is a key element in
deflagration-to-detonation transition
(DDT)~\cite{Zeldovich.et.al.-1985,Law-2006,Dorofeev-2011,Roy-et-al-review,Dorofeev-Cicarelli-review},
with two main mechanisms identified as possible causes for the flame
acceleration, namely the mechanisms of Shelkin~\cite{Shelkin} and 
Bychkov et al.~\cite{Bychkov.et.al-2008} for flame propagation in smooth and obstructed tubes respectively.
According to the Shelkin mechanism, flames accelerate in smooth
tubes due to the non-slip boundary conditions at the walls. Quantitative theory of the process has been developed and
validated by extensive  numerical simulations in
Refs.~\cite{Bychkov.et.al-2005,Akkerman.et.al-2006}. 
For the Bychkov mechanism, the delayed burning between
the obstacles in obstructed tubes produces a powerful jet flow driving
 an extremely fast flame acceleration in the unobstructed, central portion
 of the tube~\cite{Bychkov.et.al-2008,Valiev.et.al-2010}.

\begin{figure}[bp]
\includegraphics[width=0.47\textwidth\centering]{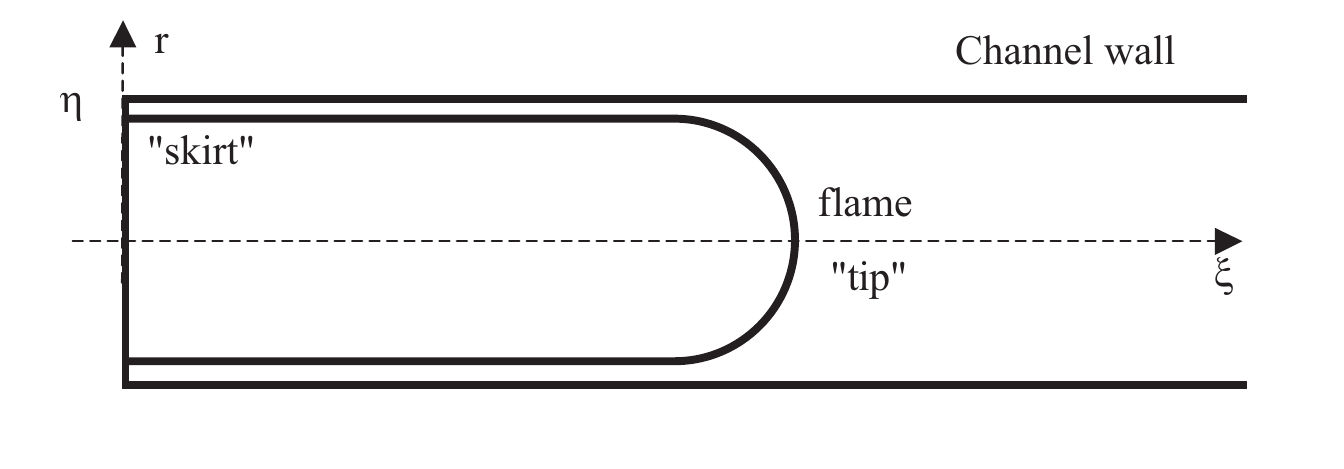}
\caption{Geometry of an accelerating finger flame.}
\label{fig-finger-sketch}
\end{figure}

 In addition to these two major mechanisms, another
possible scenario of flame acceleration in tubes  has been
demonstrated experimentally by Clanet and
Searby~\cite{Clanet.Searby-1996}. The mechanism \cite{Clanet.Searby-1996} describes the early stage 
burning of  a flame
ignited at the center of the end face of  a closed tube, leading to the transition of
an initially hemi-spherical flame kernel into a finger shaped front as illustrated in Fig.
\ref{fig-finger-sketch}. The tip of this finger flame
experiences short but powerful acceleration until the flame skirt
touches the tube walls. Thereafter, the flame acceleration stops,
the flame skirt catches up with the tip rapidly and the flame
inverts into a ``tulip'' shape. This finger
flame acceleration phenomenon was first studied in the context 
of the ``tulip flame'' formation~\cite{Clanet.Searby-1996}. However,
it was pointed out in Refs.~\cite{Bychkov.et.al-2007, Dorofeev-2011}
that the notion of a ``tulip flame'' is too ambiguous, because it may
be attributed to any concave flame front with a cusp pointing to
the burnt region, as demonstrated by several combustion phenomena of different physical 
nature~\cite{Xiao.et.al-2011,Xiao.et.al-2012,Oppenheim.Ghoniem-1983,Nkonga.et.al-1993, Pizza.et.al-2010}.
In particular, accelerating turbulent flames 
in tubes/channels
with no-slip at the walls, laminar \cite{Bychkov.et.al-2005,Akkerman.et.al-2006} or 
turbulent~\cite{Dorofeev-2011}, also exhibit an irregular ``tulip''
flame shape. Furthermore, a tulip-like flame shape is also relevant to 
oscillating flames \cite{Gonzalez.et.al-1992, Gonzalez-1996, Petchenko.et.al-2006,
 Petchenko.et.al-2007, Akkerman.et.al-2006-2, Akkerman.et.al-2010}. To avoid any ambiguity, 
 and recognizing that oscillating flames imply non-slip at the walls and relatively long flame propagation 
\cite{Akkerman.et.al-2006, Akkerman.et.al-2010}, in the
present paper we focus only on the the laminar finger flame
acceleration during the initial stage of burning, without considering
other manifestations of tulip flames. The acceleration appears to proceed in a
clearly exponential manner, as shown in a number of
works~\cite{Clanet.Searby-1996,Kuznetsov.et.al-2010,Bychkov.et.al-2007}.
This is an important effect for subsequent DDT, since
powerful precursor acceleration may create a leading shock wave
responsible for pre-heating of the fuel mixture.

A quantitative theory of finger flame acceleration in cylindrical
tubes was developed in Ref.~\cite{Bychkov.et.al-2007} by assuming incompressible flow. The theory shows that
the maximum velocity in the laboratory reference frame achieved by an 
accelerating flame tip  is
$(2\Theta-1)\Theta S_{L}$, where $S_{L}$ is the planar
unstretched flame speed and $\Theta \equiv \rho_u / \rho_b$ the
initial ratio of the fuel mixture density to the burnt gas density. For
slow hydrocarbon flames with $S_{L} \approx 40$ cm/s and
$\Theta \approx 8$, this yields a maximum velocity of the flame tip $\sim 120 S_L \sim 48$ m/s,
 which is still considerably
lower than the sound speed. The situation, however, becomes quite different
for fast hydrogen-oxygen flames, with  $S_{L} \lesssim 8$ m/s,
$\Theta \approx 8$, and the
``incompressible'' estimate of Ref.~\cite{Bychkov.et.al-2007} yields $(2\Theta-1)\Theta S_{L} \sim 960$ m/s,
which exceeds the sound speed in the
mixture, 530 m/s, and thereby fundamentally violates the incompressibility assumption. Consequently, in order to describe
finger acceleration of fast flames properly,  gas compressibility needs to be accounted for.

We next note that while the theoretical analysis
predicts exponential flame acceleration with the
incompressibility assumption~\cite{Bychkov.et.al-2005,Akkerman.et.al-2006,Bychkov.et.al-2008,Valiev.et.al-2010},
various experiments showed moderation of the initial exponential
regime with time and the possibility for the flame tip velocity to
saturate to a supersonic speed in the laboratory reference
frame~\cite{Wu.et.al-2007,Cicarelli.et.al-2010,Roy-et-al-review,Dorofeev-Cicarelli-review,Kuznetsov.et.al-2005}.
Existence of such a saturation velocity, which correlates with the
Chapman-Jouguet deflagration speed, follows from the basic theory
of deflagration and detonation
fronts~\cite{Landau.Lifshitz-1993,Chue-et-al-1993}. Recent
numerical simulation and  analytical theory demonstrated the same
tendency:  gas compressibility moderates the initial exponential
acceleration of the flame it to a slower one, 
leading to saturation of the flame
speed~\cite{Valiev.et.al-2009,Valiev.et.al-2010,Bychkov.et.al-2010a,Bychkov.et.al-2010b}.
In addition, simulations~\cite{Bychkov.et.al-2008} showed
that in obstructed channels fast flames with a relatively high
initial Mach number exhibit noticeably lower acceleration rate
as compared to slow flames. Since acceleration of finger flames
has much in common with ultra-fast flame acceleration in
obstructed channels~\cite{Bychkov.et.al-2008}, a similar influence
of gas compressibility on the finger flame acceleration is
expected.

The purpose of the present paper is to study finger flame
acceleration analytically, experimentally and computationally for
various values of the unstretched laminar flame velocity, thus
focusing on the influence of gas compressibility. The early stages of burning in tubes 
with slip adiabatic walls are considered. We first
developed an analytical theory of flame acceleration in 2D (planar)
and axisymmetric geometries through small Mach number expansion
 up to  first-order terms, demonstrating that
gas compression reduces the acceleration rate and moderates the
finger flame acceleration noticeably. We then conduct experiments
for hydrogen-oxygen mixtures with
considerably large initial values of the Mach number, showing
 the scaled
acceleration rate to be much smaller than that observed previously for
hydrocarbon flames. We also performed numerical simulations for a
wide range of initial laminar flame velocities;  the results agree well with the experiments
 as well as the theory in the limit of small gas compression
(small initial flame velocities). Similar to  previous works,
the numerical simulations show that the finger flame acceleration
is followed by the formation of a ``tulip'' flame shape, which indicates
the end of the early acceleration process.

\begin{figure}[tp]
\includegraphics[width=0.47\textwidth\centering]{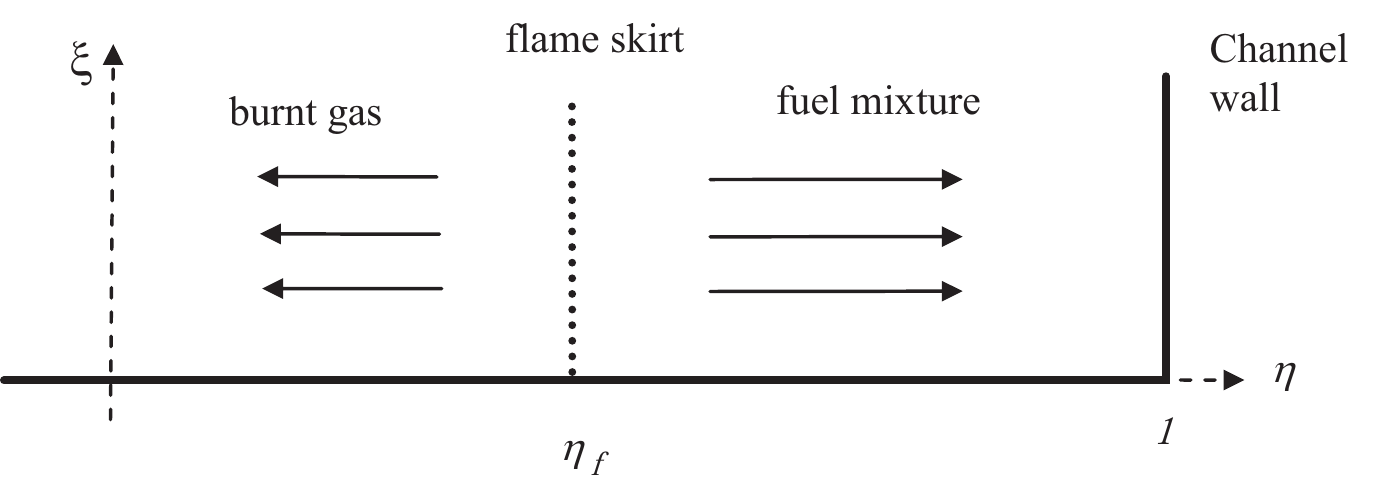}
\caption{Flow close to the channel wall.} \label{fig-wall-sketch}
\end{figure}

The paper consists of six sections. In Sections \ref{Sec.2D.incompressible}, \ref{Sec.compress.2D},
and \ref{Sec.compress.axi} we develop the  theory of flame acceleration in
the early stage of burning.  Details of the numerical
simulations are presented in Section \ref{Sec.num.method}. Section \ref{Sec.experiment} describes the
experimental results. In Section \ref{Sec.num.results} we compare
results from the theory, simulation and experiment, followed by the conclusions. Resolution and laminar velocity tests are
presented in the Appendix.


\section{Theory of finger flame acceleration}
\label{Sec.theory}

We consider a flame front propagating in a tube/channel of
radius/half-width $R$ with an ideal slip adiabatic walls as shown
in Fig.~\ref{fig-finger-sketch}. One end of the tube/channel is
kept closed, and the embryonic flame is ignited at the central
point of the closed end wall. It was explained in
Refs.~\cite{Bychkov.et.al-2007,Clanet.Searby-1996} that in the
axi-symmetric, cylindrical configuration a flame front develops
from a hemi-spherical shape at the beginning to a
finger shape. Here we first present a 2D (planar) counterpart
of this formulation, assuming flow incompressibility,
and then we extend the formulation, both in the planar and
axi-symmetric configurations, to the case of finite, but small
compressibility.








\subsection{Finger flame acceleration for planar geometry}
\label{Sec.2D.incompressible}

In the  theory we employ the standard model of an
infinitely thin flame front propagating normally with the speed
$S_{L}$, and use the dimensionless coordinates $(\eta,\xi) = (x,z)
/ R$, velocities $(w,v) = (u_{x},u_{z}) / S_{L}$, and time $\tau =
S_{L} t / R$. A 2D flame, ignited at the point $(\eta,\xi) =
(0,0)$, is initially semicircular, but the flame shape changes as
the flame-skirt $\eta _{f}$ moves along the  end wall of the chamber
($\xi = 0$) from the axis ($\eta = 0$) to the sidewall ($\eta =
1$), as shown in Fig.~\ref{fig-wall-sketch}. The flame separates
the flow into two regions of fresh mixture and burnt
gas.

\begin{figure}[tp]
\includegraphics[width=0.47\textwidth\centering]{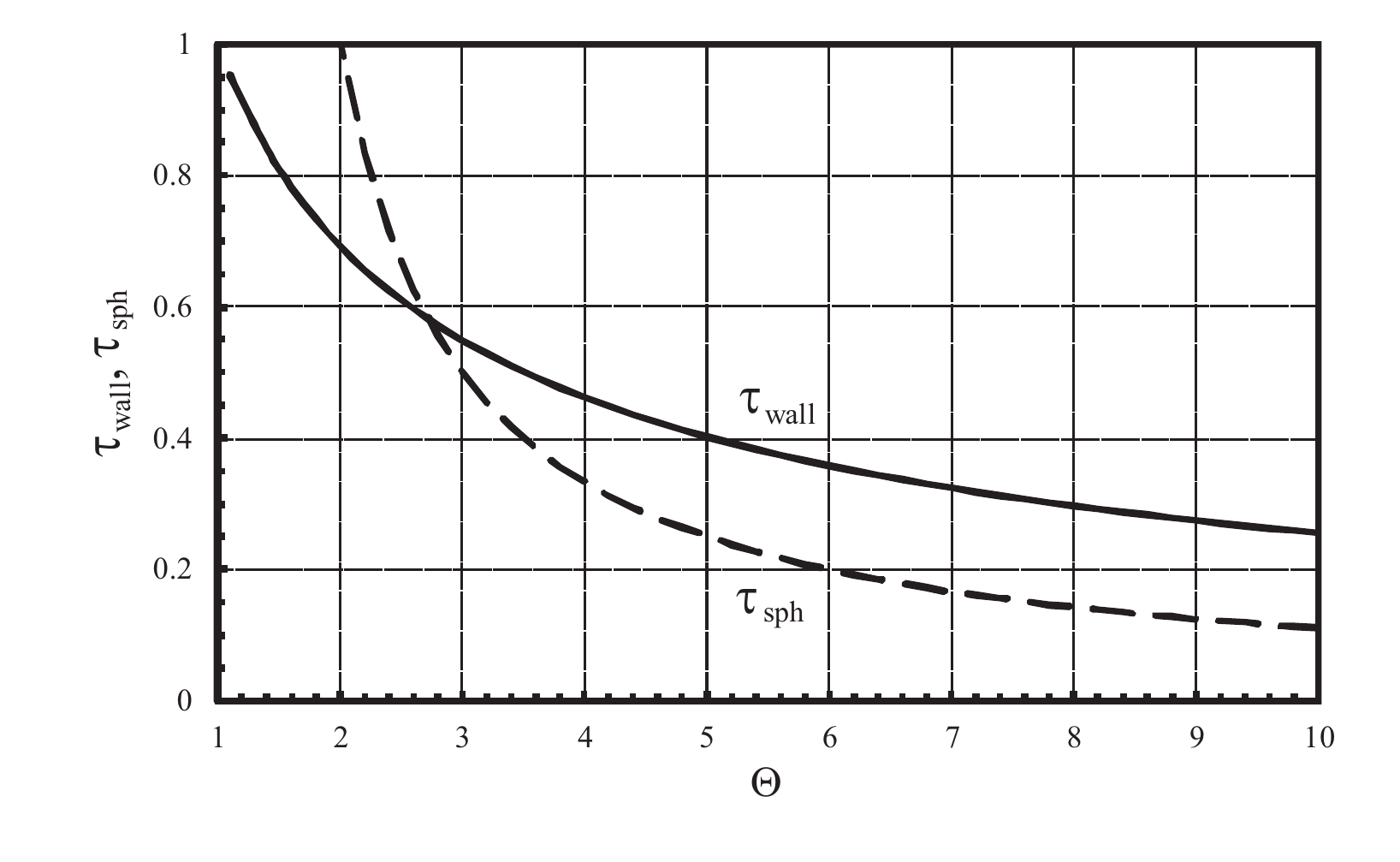}
\caption{Time limits of the flame acceleration.}
\label{fig-time-limits}
\end{figure}

Assuming incompressibility for substantially subsonic flame propagation,  the
continuity equation is given by
\begin{equation}
\label{eq4} {\frac{{\partial   w}}{{\partial \eta}} } +
{\frac{{\partial v}}{{\partial \xi}} } = 0.
\end{equation}
The boundary conditions are $v = 0$ at the end wall, $\xi = 0$, and
$w = 0$ at the side wall, $\eta = 1$. The flow in the fresh mixture
(labeled ``u'') is assumed to be potential, so Eq.~(\ref{eq4})
yields

\begin{equation}
\label{eq5}
v_{u} = C_{1} \xi , \qquad w_{u} = C_{1} \left( {1 - \eta}  \right),
\end{equation}
where the factor $C_{1}$ may depend on time, but  is independent
of the (radial) coordinate.
Recognizing that while the flow in the burnt gas (label
``b'') is rotational in general because of the curved flame shape, we can
nevertheless treat it as a potential flow close to the end wall,
where the flame front is locally planar, see
Fig.~\ref{fig-wall-sketch}. Subsequently, Eq.~(\ref{eq4}) with the
boundary condition at the channel axis, $w = 0$ at $\eta = 0$,
yields the velocity distribution in the burnt gas in the form
\begin{equation}
\label{eq7}
v_{b} = C_{2} \xi , \qquad w_{b} = - C_{2} \eta.
\end{equation}
 The matching conditions at the flame front, $\eta = \eta _{f}$,
are
\begin{equation}
\label{eq9}
{\frac{{d\eta _{f}}} {{d\tau}} } - w_{u} = 1,
\end{equation}
\begin{equation}
\label{eq10}
v_{u} = v_{b} ,
\end{equation}
\begin{equation}
\label{eq11}
w_{u} - w_{b} = \Theta - 1.
\end{equation}
Here Eq.~(\ref{eq9}) specifies the fixed propagation velocity
$S_{L}$ of the flame front with respect to the fuel mixture, while
Eqs.~(\ref{eq10}) and (\ref{eq11}) describe continuity of the
tangential velocity at the front and the jump of the normal
velocity, respectively. It is noted that the condition (\ref{eq10})
applies only at the flame skirt close to the wall. Substituting Eqs.~(\ref{eq5})--(\ref{eq7})
into Eqs.~(\ref{eq9})--(\ref{eq11}), we find $C_{1} = C_{2} =
(\Theta - 1)$. Consequently,
\begin{eqnarray}
\label{eq12-new} 
v_{u} & = & v_{b} = \left(\Theta-1\right) \xi, \nonumber \\
w_{u} & = & \left(\Theta-1\right) \left(1-\eta\right),\nonumber \\
w_{b} & = & - \left(\Theta-1\right) \eta,
\end{eqnarray}
and the evolution equation for the flame skirt, Eq.~(\ref{eq9}),
becomes
\begin{equation}
\label{eq13}
\frac{d\eta_{f}} {d\tau} - (\Theta - 1)(1 - \eta_{f} ) = 1,
\end{equation}
which can be integrated with the initial condition
$\eta_{f} = 0$ at $\tau = 0$ as
\begin{eqnarray}
\label{eq13-1} \eta _{f}  =  {\frac{{\Theta}} {{\Theta -
1}}}{\left\{ {1 - \exp {\left[ { - \left( {\Theta - 1}
\right)\tau}  \right]}} \right\}}, \nonumber \\
 \tau =  - {\frac{{1}}{{\Theta - 1}}}\ln \left( {1 -
{\frac{{\Theta - 1}}{{\Theta }}}\eta _{f}}  \right).
\end{eqnarray}
According to Eqs. (\ref{eq13}) and (\ref{eq13-1}), we identify two
 regimes of flame propagation, namely, those with the flame skirt 
close to the axis and the wall ($\eta _{f} \ll 1$ and $1 - \eta _{f} \ll 1$, respectively). 
In the limit of $\eta _{f} \ll 1$ and
$(\Theta - 1)\tau \ll 1$, the flame propagates as $d\eta _{f} /
d\tau = \Theta$, $\eta _{f} = \Theta \tau$ (i.e. $\dot {R}_{f} =
\Theta S_{L}$, $R_{f} = \Theta S_{L} t)$, which is 
related to the expansion of a semicircular flame front. In the 
limit of $1 - \eta _{f} \ll 1$, a locally planar flame ``skirt''
approaches the wall, the radial velocity of the fresh fuel mixture
tends to zero, and the flame skirt propagates with the planar
flame speed with respect to the end wall  of the channel, $d\eta _{f} /
d\tau = 1$ (i.e. $\dot {R}_{f} = S_{L})$. The time
of the transition from the hemi-circular to the ``finger''-shape
flamefront can be estimated as
\begin{equation}
\label{eq14}
\tau _{sph} \approx {\frac{{1}}{{\Theta - 1}}},
\end{equation}
when the position of the flame skirt, Eq.~(\ref{eq13-1}), is
$\eta_{f,sph} = \left(1 - e^{-1}\right) \Theta /
\left(\Theta-1\right) \approx 0.63 \, \Theta /
\left(\Theta-1\right)$
 so the transition occurs approximately when the flame
skirt has moved more than half-way to the  side wall  of the channel. Substituting $\eta_{f} = 1$ into
Eq.~(\ref{eq14}) we find the time when the flame skirt touches the tube wall
\begin{equation}
\label{eq16}
\tau _{wall} = {\frac{{\ln \Theta}} {{\Theta - 1}}}.
\end{equation}
The results~(\ref{eq14}) and (\ref{eq16}) are shown in Fig. \ref{fig-time-limits}  as functions of the expansion factor $\Theta$. It is clearly
seen from Eqs.~(\ref{eq14}) and (\ref{eq16}) and Fig. \ref{fig-time-limits} that the acceleration 
($\tau _{sph} < \tau _{wall} )$ occurs if $\Theta > e$. For $\Theta = 5\sim 10$ we have $\tau _{sph} \approx 0.11\sim 0.25$,
while $\tau _{wall} = 0.25\sim 0.4$.

\begin{figure}[tp]
\includegraphics[width=0.47\textwidth\centering]{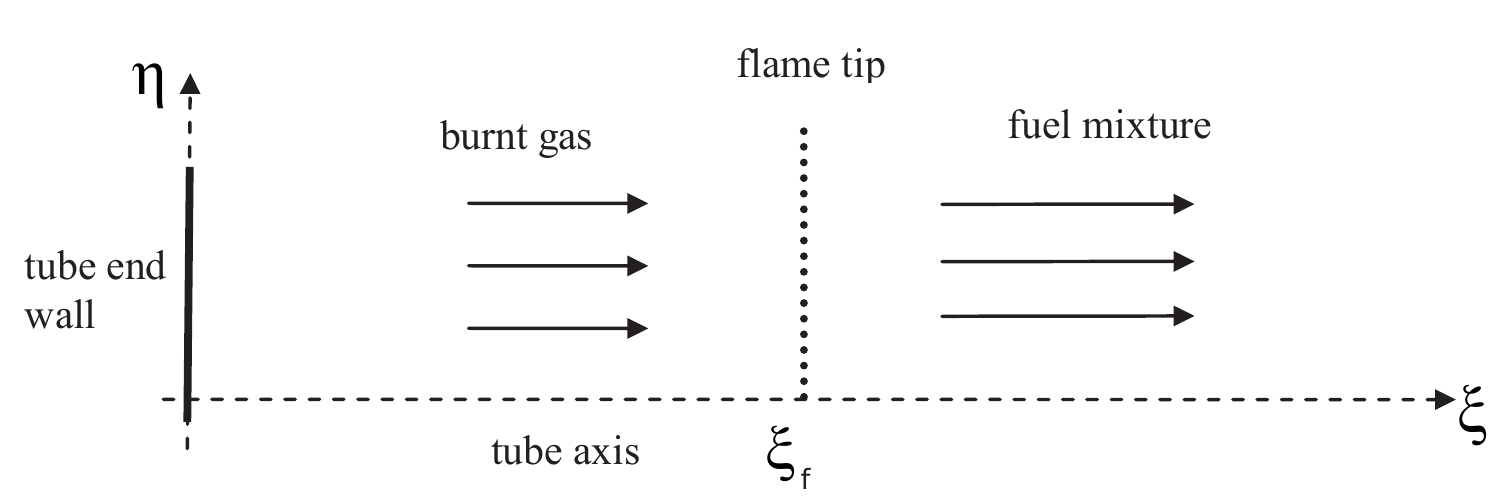}
\caption{Flow close to the channel axis.} \label{fig-axis-sketch}
\end{figure}

To determine the evolution of the flame tip, we consider the flow
along the channel axis, $\eta = 0$, as shown in
Fig.~\ref{fig-axis-sketch}. The flame shape is considered to be locally planar 
in the vicinity of the centerline ($\eta \rightarrow 0$). 
In that limit the flow can be considered to 
be potential, with the longitudinal velocity component $v$ determined by an equation similar to Eq.~(\ref{eq7}). 
The solution for the longitudinal 
velocity component $v$ has to coincide with Eq.~(\ref{eq7}) at $\eta \rightarrow 0, \ \xi \rightarrow 0$. 
Thus, the longitudinal velocity component in the burnt gas $v_b$ along the centerline ($\eta=0$) 
is governed by Eqs.~(\ref{eq7})~and~(\ref{eq12-new}). We stress that this reasoning does not 
hold away from the axis in the burnt gas, where the flow is rotational. Still, 
only the gas velocity along the centerline is utilized in the present analysis.
Based on the condition of a fixed
propagation velocity of a planar flame front with respect to the
burnt gas,
\begin{equation}
\label{eq17}
\frac{d\xi_{tip}} {d\tau} - v_{b} = \Theta,
\end{equation}
with $v_{b}$ given by Eq.~(\ref{eq12-new}), we arrive at the
differential equation for the flame tip,
\begin{equation}
\label{eq18}
\frac{d\xi _{tip}} {d\tau} - (\Theta - 1)\xi_{tip} = \Theta,
\end{equation}
with the initial condition $\xi_{tip} \left(0\right) = 0$, and the solution
\begin{equation}
\label{eq19}
\xi_{tip} = {\frac{{\Theta}} {{\Theta - 1}}}{\left\{ {\exp {\left[ {\left(
{\Theta - 1} \right) \tau} \right]} - 1} \right\}},
\end{equation}
which yields a semicircular flame front just after ignition,
$(\Theta - 1)\tau \ll 1$, when Eq.~(\ref{eq19}) is reduced to
$\xi_{tip} = \Theta \tau = \eta _{f}$, and the transition to the
exponential acceleration thereafter.  According to Eq. (\ref{eq19}),
the growth rate during the exponential stage of acceleration for the planar geometry is given by
\begin{equation}
\label{eq19-1}
\sigma_{0,pl} = \Theta-1.
\end{equation}

\noindent  At the end of the
acceleration, when the flame skirt touches the wall, we have $\tau
= \tau_{wall}$, so the position of the flame tip is
\begin{equation}
\label{eq20}
\xi_{wall} = \xi_{tip} \left( \tau_{wall} \right) = \Theta,
\end{equation}
or $Z_{wall} = \Theta R$ in  dimensional units. Therefore, accounting for Eq.~(\ref{eq18}), we have
\begin{equation}
\label{eq201} \left( \frac{d \xi_{tip}}{d \tau} \right)_{wall} =
\left( \frac{u _{tip}}{S_{L}} \right)_{max} = \Theta^2 .
\end{equation}

To describe the flame shape and evaluate the total increase in the flame surface area during the flame acceleration, 
we assume, realistically, that at
 the end of the acceleration the flame shape is almost self-similar, with $\eta _{f} \approx 1$ and
$\xi _{tip} \propto \exp {\left[ {\left( {\Theta - 1} \right)\tau}  \right]} \gg 1$. We therefore look for the flame shape
in the form
\begin{eqnarray}
\label{eq21} 
\xi_{f} & = & \xi _{tip} (\tau ) - f(\tau ,\eta ), \nonumber \\
f(\tau ,\eta ) & = & \varphi \left( {\eta}  \right)\exp {\left[
{\left( {\Theta - 1} \right)\tau}  \right]}.
\end{eqnarray}
The accuracy of such an approximation is $\Theta^{-1} \ll 1$, which is acceptable for typical flames. Then the equation
of flame evolution with respect to the burnt gas can be written as
\begin{equation}
\label{eq22}
{\frac{{\partial f}}{{\partial \tau}} } + w_{b} {\frac{{\partial f}}
{{\partial \eta}} } - v_{b} = \Theta {\left[ {1 + \left(
{{\frac{{\partial f}}{{\partial \eta}} }} \right)^{2}} \right]}^{1/2}
\approx \Theta {\frac{{\partial f}}{{\partial \eta}} }.
\end{equation}
Accounting for the exponential state of flame acceleration,
Eq.~(\ref{eq21}) and the velocity distribution~(\ref{eq7}), we
reduce Eq.~(\ref{eq22}) to
\begin{equation}
\label{eq23}
{\left[ {\Theta - (\Theta - 1)\eta}\right]}{\frac{{d\varphi}} {{d\eta}}} = \Theta,
\end{equation}
with the boundary condition at the axis, $\varphi \left(0\right) = 0$, yielding the solution
\begin{equation}
\label{eq24}
\varphi \left( {\eta}  \right) = - {\frac{{\Theta}} {{\Theta - 1}}}\ln
\left( {1 - {\frac{{\Theta - 1}}{{\Theta}} }\eta}  \right),
\end{equation}
\noindent with
\begin{eqnarray}
\label{eq25} \xi_f \left( \eta ,\tau  \right)  = \frac{\Theta} {\Theta - 1} \{ \left[ {1 - \ln \left(
{1 - \frac{\Theta - 1}{\Theta}\eta} \right)} \right] \times \nonumber \\
 \times \exp {\left[ {\left( {\Theta - 1} \right)\tau} \right]} - 1 \}.
\end{eqnarray}
Then the maximum increase in the flame length, achieved when the flame skirt touches the wall, is
\begin{eqnarray}
\label{eq26} L_{w} / 2R  = {\int\limits_{0}^{1} {\sqrt {1 + \left(
{{\frac{{\partial f}} {{\partial \eta}} }} \right)^{2}}}} d\eta
\approx {\int\limits_{0}^{1} {{\frac{{\partial f}}{{\partial
\eta}} }}} d\eta =  \nonumber \\
= f(1,\tau _{wall} ) =  \varphi
\left( {1} \right)\exp {\left[ {\left( {\Theta - 1} \right)
\tau_{wall}} \right]} = {\frac{{\Theta ^{2}\ln \Theta}} {{\Theta -
1}}}. \nonumber \\
\end{eqnarray}

\subsection{Influence of gas compressibility on acceleration rate for planar geometry}
\label{Sec.compress.2D}

We  next consider the problem, with 
first-order accuracy for the initial flame propagation Mach number $Ma
\equiv S_{L} / c_{S,0} \ll 1$, where $c_{S,0}$ is the initial
sound speed of the fresh mixture. This approach is conceptually
close to that of
Refs.~\cite{Bychkov.et.al-2010a, Bychkov.et.al-2010b}. The
compressible counterpart of Eq.~(\ref{eq17}) for the dynamics of
the flame tip becomes
\begin{equation}
\label{eq27} {\frac{{d\xi _{tip}}} {{d\tau}} } - v_{bf} =
\vartheta,
\end{equation}
where $v_{bf} = v_{b} (\xi _{f} ,\tau )$ is the flow velocity of the burnt gas in the $\xi$-direction just at
the flame front, and $\vartheta \equiv \rho _{u} / \rho _{bf}$ the instantaneous gas expansion factor, with
$\vartheta _{0} \equiv \Theta$.

As long as $Ma \ll 1$, we can treat the flow ahead of the flame as isentropic. In this case, to first order
in $Ma$, we have the following relations for the scaled density $\tilde {\rho} _{u} = \rho _{u} / \rho _{u,0} $,
pressure $\tilde {P}_{u} = P_{u} / P_{u,0} $, and temperature $\tilde {T}_{u} = T_{u} / T_{u,0} $ in the fuel mixture
\begin{eqnarray}
\label{eq28} 
\tilde {\rho} _{u} & = & \left( {1 + {\frac{{\gamma - 1}} {{2}}}{\frac{{v_{u} }}{{c_{S, 0}}} }}
\right)^{{\frac{{2}}{{\gamma - 1}}}} \approx 1 + {\frac{{v_{u}}} {{c_{S, 0}}} }  \approx \nonumber \\ 
& \approx &  1 + Ma(\Theta - 1) \left( {\xi {}_{tip} +
1} \right),
\end{eqnarray}
\begin{eqnarray}
\label{eq29} 
\tilde {P}_{u} & = & \left( {1 + {\frac{{\gamma - 1}}
{{2}}}{\frac{{v_{u} }}{{c_{S, 0}}} }}\right)^{{\frac{{2\gamma}}
{{\gamma - 1}}}} \approx 1 + \gamma {\frac{{v_{u}}} {{c_{S, 0}}} }   \approx\nonumber \\
& \approx &   1 + Ma\gamma (\Theta - 1)\left( {\xi _{tip} + 1} \right),
\end{eqnarray}
\begin{eqnarray}
\label{eq30} \tilde {T}_{u} & = & \left( {1 + {\frac{{\gamma -
1}}{{2}}} {\frac{{v_{u} }}{{c_{S, 0}}} }} \right)^{2} \approx 1 +
(\gamma - 1){\frac{{v_{u}}} {{c_{S, 0} }}}  \approx \nonumber \\ & \approx & 1 + Ma(\gamma
- 1) (\Theta - 1)\left( {\xi _{tip} + 1} \right),
\end{eqnarray}
where $\gamma \equiv C_{P} / C_{V} $ is the adiabatic index. It is
noted that Eqs.~(\ref{eq28})--(\ref{eq30}) also define the
rigorous mathematical limit of validity for the present theory,
i.e. $v_{u} / c_{S, 0} \ll 1$. The matching relations at the flame
front are
\begin{equation}
\label{eq31}
\tilde {\rho} _{u} \left( {{\frac{{d\xi _{tip}}} {{d\tau}} } - v_{u}}
\right) = \tilde {\rho} _{bf} \left( {{\frac{{d\xi _{tip}}} {{d\tau}} } -
v_{bf}} \right),
\end{equation}
\begin{equation}
\label{eq32}
\tilde {P}_{u} + \tilde {\rho} _{u} \left( {{\frac{{d\xi _{tip}}} {{d\tau }}}
- v_{u}}  \right)^{2} = \tilde {P}_{bf} + \tilde {\rho} _{bf} \left(
{{\frac{{d\xi _{tip}}} {{d\tau}} } - v_{bf}}  \right)^{2},
\end{equation}
\begin{eqnarray}
\label{eq33} \tilde {T}_{u}  +  \tilde {Q} + {\frac{{1}}{{2C_{P}
T_{u,0}}} }\left( {{\frac{{d\xi _{tip}}} {{d\tau}} } - v_{u}}
\right)^{2} = \nonumber \\ = \tilde {T}_{bf} + {\frac{{1}}{{2C_{P} T_{u,0}}}
}\left( {{\frac{{d\xi _{tip}}} {{d\tau}} } - v_{bf}} \right)^{2},
\end{eqnarray}
where $\tilde {Q} \equiv Q / C_{P} T_{u,0} = \Theta - 1$ is the scaled reaction heat release. With the first-order
approximation for small $Ma$, we reduce Eqs.~(\ref{eq32}) and (\ref{eq33}) to
\begin{equation}
\label{eq34} \tilde {P}_{u} = \tilde {P}_{bf} ,\ \tilde {T}_{u} +
\Theta - 1 = \tilde {T}_{bf} .
\end{equation}
Using the perfect gas law, $\tilde {\rho} _{u} \tilde {T}_{u} = \tilde {\rho} _{bf} \tilde {T}_{bf}$,
as well as Eq.~(\ref{eq30}), we find
\begin{equation}
\label{eq35}
\vartheta = 1 + {\frac{{\Theta - 1}}{{\tilde {T}_{u}}} } = \Theta -
Ma(\gamma - 1)(\Theta - 1)^{2}\left( {\xi _{tip} + 1} \right).
\end{equation}
While, according to the Euler equation
\begin{equation}
\label{eq36}
{\frac{{\partial v_{b}}} {{\partial \tau}} } + v_{b} {\frac{{\partial v_{b}
}}{{\partial \xi}} } = - {\frac{{1}}{{\tilde {\rho} _{b}}} }{\frac{{\partial
\tilde {P}_{b}}} {{\partial \xi}} },
\end{equation}
pressure is uniform in the burnt gas, $\tilde {P}_{b} (\tau) = \tilde {P}_{bf} $, up to the first-order in $Ma \ll 1$,
it however grows in time, and thereby increases the temperature and density of the burnt gas due to adiabatic compression.

We next consider propagation of the nearly planar flame ``skirt''. Within the accuracy of \textit{Ma}, the pressure
in the fuel mixture between the flame front and the sidewall is the same as that in the burnt gas, $\tilde {P}_{s} = \tilde
{P}_{b} (\tau ) = \tilde {P}_{bf} = \tilde {P}_{u} $. Thus the density and temperature of the fuel
mixture around the flame skirt are the same as those of the fuel mixture just ahead of the flame tip, since in both cases
we have adiabatic compression and the same final pressure. Consequently, the continuity equation for the fuel in the
domain between the flame shirt and the side wall, $\eta _{f} < \eta < 1$, takes the form
\begin{equation}
\label{eq37} {\frac{{\partial w_{s}}} {{\partial \eta}} } \approx
- {\frac{{1}}{{\tilde {\rho} _{s}}} }{\frac{{d\tilde {\rho} _{s}}}
{{d\tau}} } = - {\frac{{1}}{{\gamma   \tilde {P}_{s}}}
}{\frac{{d\tilde {P}_{s} }}{{d\tau}} } = - {\frac{{1}}{{\gamma
\tilde {P}_{u} }}}{\frac{{d\tilde {P}_{u}}} {{d\tau}}},
\end{equation}
with the following solution satisfying the matching relation at the side wall, $\eta = 1$, $w_{s} = 0$,
\begin{eqnarray}
\label{eq38}
w_{s} &  = & {\frac{{1}}{{\gamma   \tilde {P}_{u}}}
}{\frac{{d\tilde {P}_{u}}} {{d\tau}} }\left( {1 - \eta}  \right), \nonumber \\
 w_{s,f} & = & {\frac{{1}}{{\gamma   \tilde {P}_{u}}}}
{\frac{{d\tilde {P}_{u}}} {{d\tau}} }\left( {1 - \eta _{f}}
\right).
\end{eqnarray}
Here the subscript ``$s,f$'' designates the flow velocity just ahead of the flame skirt. Using the matching relations at the
flame front, we find the velocity in the burnt gas at the flame skirt, subscripted by ``$s,b,f$'', as
\begin{equation}
\label{eq39} w_{s,b,f} = w_{s,f} - \vartheta + 1 =
{\frac{{1}}{{\gamma   \tilde {P}_{u}}} }{\frac{{d\tilde {P}_{u}}}
{{d\tau}} }\left( {1 - \eta _{f}} \right) - \vartheta + 1.
\end{equation}

We subsequently solve the continuity equation, equivalent to Eq.~(\ref{eq37}), in the burnt gas with Eq.~(\ref{eq39}) to
find
\begin{eqnarray}
\label{eq40} w_{b} & = & {\frac{{1}}{{\gamma  \tilde {P}_{u}}}
}{\frac{{d\tilde {P}_{u}}} {{d\tau}} }\left( {1 - \eta _{f}}
\right) - \vartheta + 1 + {\frac{{1}}{{\gamma  \tilde {P}_{u}}}
}{\frac{{d\tilde {P}_{u} }}{{d\tau}} }\left( {\eta _{f} - \eta}
\right) = \nonumber \\ & = & {\frac{{1}}{{\gamma   \tilde {P}_{u}}}
}{\frac{{d\tilde {P}_{u}}} {{d\tau}} }\left( {1 - \eta } \right) -
\vartheta + 1.
\end{eqnarray}
Similarly, the continuity equation for the burnt gas around the
symmetry axis takes the form
\begin{equation}
\label{eq41} {\frac{{\partial w_{b}}} {{\partial \eta}} } +
{\frac{{\partial v_{b} }}{{\partial \xi}} } = -
{\frac{{1}}{{\gamma   \tilde {P}_{u} }}}{\frac{{d\tilde {P}_{u}}}
{{d\tau}} },
\end{equation}
which can be integrated as
\begin{equation}
\label{eq42} v_{b} = \left( {\vartheta - 1} \right)\xi -
{\frac{{1}}{{\gamma \tilde {P}_{u}}} }{\frac{{d\tilde {P}_{u}}}
{{d\tau}} }\xi , \qquad w_{b} = - \left( {\vartheta - 1}
\right)\eta .
\end{equation}
Consequently, the evolution equation for the flame tip, Eq.~(\ref{eq27}), becomes
\begin{equation}
\label{eq43} {\frac{{d\xi _{tip}}} {{d\tau}} } = \left( {\vartheta - 1 -
{\frac{{1}}{{\gamma \tilde {P}_{u}}}}{\frac{{d\tilde
{P}_{u} }}{{d\tau}} }} \right)\xi _{tip} + \vartheta.
\end{equation}

Finally, substituting Eqs.~(\ref{eq29}) and (\ref{eq35}) into Eq.~(\ref{eq43}), neglecting the second- and higher-order
terms in $Ma$, and accounting for the zeroth-order approximation, Eq.~(\ref{eq19}), we obtain
\begin{equation}
\label{eq44} {\frac{{d\xi _{tip}}} {{d\tau}} } = - Ma\gamma
\left( {\Theta - 1} \right)^{2}\xi _{tip}^{2} + \sigma _{1,pl} \xi_{tip}
+ \Theta_{1},
\end{equation}
with
\begin{equation}
\label{eq45-1} \sigma _{1,pl} = \left( {\Theta - 1} \right){\left[ {1
- Ma\left( {\Theta + 2\left( {\gamma - 1} \right)\left( {\Theta - 1} \right)}
 \right)} \right]},
\end{equation}
\begin{equation}
\label{eq45} \Theta_{1} = \Theta - Ma(\gamma - 1)(\Theta - 1)^{2}.
\end{equation}
In the limit of incompressible flow, $Ma = 0$, we have $\sigma _{1,pl} = \Theta - 1$,
$\Theta_{1} = \Theta$, and Eq.~(\ref{eq44}) fully reproduces Eq.~(\ref{eq18}).
Accounting for gas compressibility, we obtain moderation of the flame acceleration
in Eq.~(\ref{eq44}), which is described by two types of terms: linear and nonlinear
with respect to $\xi_{tip}$. The linear term does not change the exponential state
of the flame acceleration, though they reduce the acceleration rate to $\sigma_{1,pl}$
as compared to $\Theta - 1$ for the incompressible flow. At the very beginning,
for $\xi _{tip} \to 0$, the flame acceleration is moderated by the linear terms
only, and Eq.~(\ref{eq44}) reduces to
\begin{equation}
\label{eq46} \frac{d\xi_{tip}}{d\tau} =
\sigma _{1,pl} \xi_{tip} + \Theta_{1},
\end{equation}
with the solution
\begin{equation}
\label{eq47} \xi _{tip} = {\frac{{\Theta _{1}  }} {{\sigma _{1,pl}}}}
{\left[ {\exp \left( {\sigma _{1,pl} \tau}  \right) - 1} \right]}.
\end{equation}
The nonlinear term of Eq.~(\ref{eq44}), however, becomes important quite fast and modifies the exponential state
of flame acceleration to a slower one. The complete analytical solution to Eq.~(\ref{eq44}) is given by
\begin{equation}
\label{eq48}
\xi_{tip} = {\frac{{2\Theta _{1} \,{\left[ {\exp \left( {\sigma _{2} \tau}
\right) - 1} \right]}}}{{\left( {\sigma _{2} - \sigma _{1,pl}}  \right)\exp
\left( {\sigma _{2} \tau}  \right) + \left( {\sigma _{2} + \sigma _{1,pl}}
\right)}}} ,
\end{equation}
where $\sigma_{2} \equiv \sqrt {\sigma_{1,pl}^{2} + 4 Ma\gamma \Theta_{1} \left( \Theta - 1 \right)^{2}}$.

\subsection{Influence of gas compressibility for the axi-symmetric geometry}
\label{Sec.compress.axi}

Now we reconsider the problem in an
axi-symmetric, cylindrical tube, Ref.~\cite{Bychkov.et.al-2007},
incorporating compressibility with the accuracy of the first order
for flame propagation Mach number.

The
incompressible continuity equation in the axi-symmetric geometry
takes the form~\cite{Bychkov.et.al-2007}
\begin{equation}
\label{eq4-axi}
\frac{1}{\eta} \frac{\partial}{\partial \eta}
\left(\eta w\right) + \frac{\partial v}{\partial \xi} = 0,
\end{equation}
and the axi-symmetric counterparts for the velocity distributions,
Eq.~(\ref{eq12-new}), and the evolution of the flame skirt,
Eqs.~(\ref{eq13})--(\ref{eq13-1}), are given by
\begin{eqnarray}
\label{eq12-axi} v_{u} & = & v_{b} = 2 \left(\Theta-1\right)
\eta_{f} \xi,\nonumber \\
w_{u} & = & \left(\Theta - 1\right) \eta_{f}
\left(\frac{1}{\eta} - \eta\right), \nonumber \\
w_{b} & = & -\left(\Theta-1\right) \eta_{f} \eta,
\end{eqnarray}
\begin{equation}
\label{eq13-axi}
\frac{d\eta_{f}} {d\tau} - (\Theta - 1)(1 - \eta_{f}^{2} ) = 1
\qquad \Longrightarrow
\end{equation}
\begin{eqnarray}
\label{eq63} \eta_{f} & = & \frac{\Theta}{\alpha}\tanh \left(\alpha
\tau \right), \nonumber \\
 \tau_{wall} & = &
\frac{1}{2\alpha} \ln \left( \frac{\Theta + \alpha} {\Theta -
\alpha}\right), \nonumber \\
 \tau_{sph} &  = & \frac{1}{2\alpha},
\end{eqnarray}
where
\begin{equation}
\label{eq-alpha}
\alpha = \sqrt{\Theta \left(\Theta - 1\right)}.
\end{equation}

It can be readily shown from Eq. (\ref{eq63}) that acceleration is possible
(i.e. $\tau_{sph}<\tau_{wall}$) if $\Theta > \left({1-[(e-1)/e+1)]^2}\right)^{-1} \approx 1.27$. For $\Theta = 5 \sim 10$,
Eq. (\ref{eq63}) yields $\tau_{sph} \approx 0.05 \sim 0.11$, and $\tau_{wall} \approx 0.19 \sim 0.32$.
These quantities are much smaller than those of the planar geometry, Eqs. (\ref{eq14}) and (\ref{eq16}),
making an indirect proof that for the axisymmetric geometry acceleration proceeds faster than the planar one.

With the result~(\ref{eq12-axi}), equation for the flame tip,
$\dot{\xi_{tip}}-v_{b}=\Theta$, becomes
\begin{equation}
\label{2.3.eq20} \frac{d\xi_{tip}} {d\tau} - 2(\Theta - 1) \eta_{f} (\tau)
\xi_{tip}(\tau ) = \Theta,
\end{equation}
or
\begin{equation}
\label{2.3.eq20-I} \frac{d\xi_{tip}} {d\tau} -
2 \alpha \tanh \left( \alpha \tau \right) \xi_{tip} = \Theta,
\end{equation}
with the solution
\begin{equation}
\label{2.3.eq20-II}
\xi_{tip} = \frac{\Theta}{2\alpha} \sinh \left(2 \alpha \tau \right),
\end{equation}
which also yields $\xi_{wall} = \Theta$ similar to the 2D
result~(\ref{eq20}). At sufficiently late times we have
$\eta_{f}\approx 1$ and Eqs.~(\ref{2.3.eq20})--(\ref{2.3.eq20-I})
reduce to
\begin{eqnarray}
\label{2.3.eq23}
\frac{d\xi_{tip}} {d\tau} = 2 \alpha \xi_{tip} + \Theta \qquad \Longrightarrow \nonumber \\
\qquad \xi_{tip} = \frac{\Theta}{2\alpha}
\left[ \exp \left( 2 \alpha \tau \right) - 1 \right],
\end{eqnarray}
so the flame tip accelerates almost exponentially,
with the acceleration rate
\begin{equation}
\label{2.3.eq23-1}
\sigma_{0,axi} = 2 \alpha = 2\sqrt{\Theta(\Theta - 1)}.
\end{equation}
This result exceeds considerably (by a factor of about 2) its 2D
counterpart ($\Theta-1$), see Eq.~(\ref{eq19-1}), and it is 
slightly smaller than the model estimation $2 \Theta$ of Clanet and
Searby~\cite{Clanet.Searby-1996}.

Now we account for small, but finite gas compression. With
the axial velocity $v_{u}$ given by Eq.~(\ref{eq12-axi}), the
axi-symmetric counterparts of Eqs.~(\ref{eq28})--(\ref{eq30}) and
(\ref{eq35}) are
\begin{eqnarray}
\label{2.3.eq35}
\tilde{\rho}_{u} & = & \left(1 + \frac{\gamma - 1}{2} \frac{v_{u}}{c_{0}}
\right)^{\frac{2}{\gamma - 1}} \approx \nonumber \\
& \approx & 1 + Ma\left( 2\alpha \xi_{tip} + \Theta - 1 \right),
\end{eqnarray}
\begin{eqnarray}
\label{2.3.eq36}
\tilde{P}_{u} & = & \left(1 + \frac{\gamma - 1}{2} \frac{v_{u}}{c_{0}}
\right)^{\frac{2\gamma}{\gamma - 1}} \approx \nonumber \\
& \approx & 1 + Ma\gamma \left( 2\alpha \xi_{tip} + \Theta - 1 \right),
\end{eqnarray}
\begin{eqnarray}
\label{2.3.eq37}
\tilde {T}_{u} & = & \left( 1 + \frac{\gamma-1}{2} \frac{v_{u}}{c_{0}} \right)^{2} \approx \nonumber \\
& \approx & 1 + Ma \left(\gamma-1\right) \left(2 \alpha \xi_{tip} +
\Theta - 1 \right),
\end{eqnarray}
\begin{equation}
\label{2.3.eq42} \vartheta = 
\Theta - Ma(\gamma - 1)(\Theta - 1)^{2} \left(2 \frac{\Theta}{\alpha}
\xi_{tip} + 1 \right).
\end{equation}
Following the strategy of Section \ref{Sec.compress.2D}, we find
\begin{equation}
\label{eq59} 
{\frac{{1}}{{\eta}} }{\frac{{\partial}} {{\partial
\eta}} }\left( {\eta w_{b}}  \right) + {\frac{{\partial v_{b}}}
{{\partial \xi}} } = - {\frac{{1}}{{\gamma   \tilde {P}_{u}}}}
\frac{{d\tilde {P}_{u} }}{{d\tau}} \qquad \Longrightarrow \qquad
\end{equation}
\begin{eqnarray}
\label{eq60}
v_{b} & = & 2\left( {\vartheta - 1} \right)\eta _{f} \xi - {\frac{{1}}{{\gamma
\tilde {P}_{u}}} }{\frac{{d\tilde {P}_{u}}} {{d\tau}} }\xi {\rm,} \nonumber \\
w_{b} & = & - \left( {\vartheta - 1} \right) \eta_{f} \eta.
\end{eqnarray}
Similar to Eqs.~(\ref{eq13-axi}) and (\ref{eq63}), the flame skirt
position is given by
\begin{equation}
\label{eq63-new}
\frac{d\eta_{f}} {d\tau} - (\vartheta - 1)(1 - \eta_{f}^{2} ) = 1
\qquad \Longrightarrow \qquad
\eta_{f} = \frac{\vartheta}{\hat{\alpha}} \tanh \left( \hat{\alpha} \tau \right),
\end{equation}
where $\hat{\alpha} = \sqrt{\vartheta \left(\vartheta -
1\right)}$,
and the evolution equation for the flame tip is
\begin{equation}
\label{eq62} {\frac{{d\xi _{tip}}} {{d\tau}} } = \left( {2\left(
{\vartheta - 1} \right)\eta _{f} - {\frac{{1}}{{\gamma \tilde
{P}_{u} }}}{\frac{{d\tilde {P}_{u}}} {{d\tau}} }} \right) \xi_{tip} + \vartheta.
\end{equation}
Holding the zeroth- and first-order approximations for $Ma$ in Eq.~(\ref{eq62}), we rewrite it in the form
\begin{eqnarray}
\label{eq62_I} \frac{d\xi _{tip}}{d\tau}  =  \left[2 \hat{\alpha}
\tanh \left( \hat{\alpha} \tau \right) - 2 Ma \vartheta \alpha
\right] \xi_{tip} - \nonumber \\
-  4 Ma  \alpha  \hat{\alpha} \tanh \left(
\hat{\alpha} \tau \right) \xi_{tip}^{2} + \vartheta,
\end{eqnarray}
or
\begin{eqnarray}
& & \label{eq62_II} \frac{d\xi _{tip}}{d\tau} = \left\{2 \alpha_{1}
\tanh \left( \alpha_{1} \tau \right) - 2 Ma \, \alpha \left[1 +
\gamma \left(\Theta - 1\right) \right] \right\} \xi_{tip} -
\nonumber \\
& & - 2 Ma \left\{ \frac{\alpha \tau B}{\cosh^{2} \left( \alpha \tau
\right)} + \left( 2 \alpha^{2} - B \right) \tanh \left( \alpha
\tau \right) \right\} \xi^{2}_{tip}  + \Theta_1 \nonumber \\
\end{eqnarray}
where $\Theta_1$ is the same as in the planar geometry, see
Eq.~(\ref{eq45}),
and
\begin{eqnarray}
\label{eq_alpha_1} \alpha_{1} & = & \sqrt{\Theta_{1}
\left(\Theta_{1} - 1\right)},
\nonumber \\
B & = & \left(\gamma - 1 \right) \left(\Theta - 1 \right)
\left(2 \Theta - 1 \right).
\end{eqnarray}

In general, Eq.~(\ref{eq62_II}) has to be solved computationally,
but we shall integrate it analytically with several asymptotic
approaches. First, in the limit of $Ma = 0$, we have
$\Theta_{1} = \Theta$, $\alpha_{1} = \alpha$ and
Eq.~(\ref{eq62_II}) reproduces Eq.~(\ref{2.3.eq20-I}). Even for finite gas compressibility,
the effect of the nonlinear term in Eq.~(\ref{eq62_II}) is
negligible in the very beginning, hence Eq.~(\ref{eq62_II}) can be
approximated by
\begin{eqnarray}
\label{eq62_III} \frac{d\xi _{tip}}{d\tau} & = &
\{2 \alpha_{1} \tanh \left( \alpha_{1} \tau \right) - \nonumber \\
& - & 2 Ma \, \alpha \left[1 + \gamma \left(\Theta - 1\right) \right] \}
\xi_{tip} + \Theta_{1},
\end{eqnarray}
with the solution to the first-order approximation for $Ma$ being
\begin{eqnarray}
\label{eq62_IV} 
\xi_{tip} &  = & \frac{\Theta_1}{2\alpha_1} \sinh
\left( 2 \alpha_{1} \tau \right) - \nonumber \\
& - & 2  Ma   \frac{\Theta}{\alpha}
\left[1 + \gamma \left(\Theta - 1\right) \right] \cosh^2
\left(\alpha \tau \right) \ln \cosh \left( \alpha \tau \right). \nonumber \\
\end{eqnarray}
The nonlinear term of Eq.~(\ref{eq62_II}), however, becomes
important quite fast, modifying the state of flame acceleration to
a slower one, and hence making the asymptote~(\ref{eq62_IV})
incorrect. However, at a sufficiently late stage of the
acceleration, we can approximate $\tanh \left( \alpha_{1} \tau
\right) \sim 1$, hence $\eta_{f} \sim \vartheta / \hat{\alpha}$,
and Eq.~(\ref{eq62_II}) is reduced to a form similar to that of the 2D,
Eq.~(\ref{eq44}) ,
\begin{equation}
\label{eq67}
{\frac{{d\xi _{tip}}} {{d\tau}} } = - Ma \psi \xi_{tip}^{2} +
\sigma_{1,axi} \xi _{tip} + \Theta _{1},
\end{equation}
where
\begin{equation}
\label{eq68}
\psi = 2\left( {\Theta - 1} \right)
\left(2\Theta \gamma - \gamma + 1\right),
\end{equation}
\begin{equation}
\label{eq69}
\sigma_{1,axi} = \sigma _{0,axi} {\left\{ {1 -
Ma{\left[ {\Theta +  \frac{4 \Theta -1 }{2\Theta} 
\left( {\gamma - 1} \right)\left( {\Theta - 1} \right)}
\right]}} \right\}},
\end{equation}
and with the solution, Eq.~(\ref{eq48}),
\begin{eqnarray}
\label{eq74}
\xi_{tip} = {\frac{{2\Theta_{1}
{\left[ {\exp \left( {\sigma _{2} \tau} \right) - 1}
\right]}}}{{\left( {\sigma _{2} - \sigma _{1,axi}}  \right)\exp \left(
{\sigma _{2} \tau}  \right) + \left( {\sigma _{2} + \sigma _{1,axi}}
\right)}}},
\end{eqnarray}
\noindent
 where
$\sigma_{2} \equiv \sqrt {\sigma_{1,axi}^{2} + 4 \Theta_{1} Ma\psi}$  in the axisymmetric configuration.
Obviously, the result~(\ref{eq67}) to (\ref{eq74}) fully recovers
the properties of its planar counterpart: in the limit of $Ma = 0$
we have $\sigma_{1,axi} = \sigma _{0,axi}$, and Eqs.~(\ref{eq67}) and
(\ref{eq74}) reduce to Eq.~(\ref{2.3.eq23}); accounting for gas
compressibility, we obtain linear and nonlinear moderation of the
flame acceleration with respect to $\xi_{tip}$. Reducing the
acceleration rate from $\sigma_{0,axi}$ to $\sigma_{1,axi}$, the linear
terms do not change the exponential state of acceleration while
the nonlinear term of Eq.~(\ref{eq67}) modifies the exponential
state of flame acceleration to a slower one as soon as it becomes
important.

\begin{figure}[tp]
\includegraphics[width=0.49\textwidth\centering]{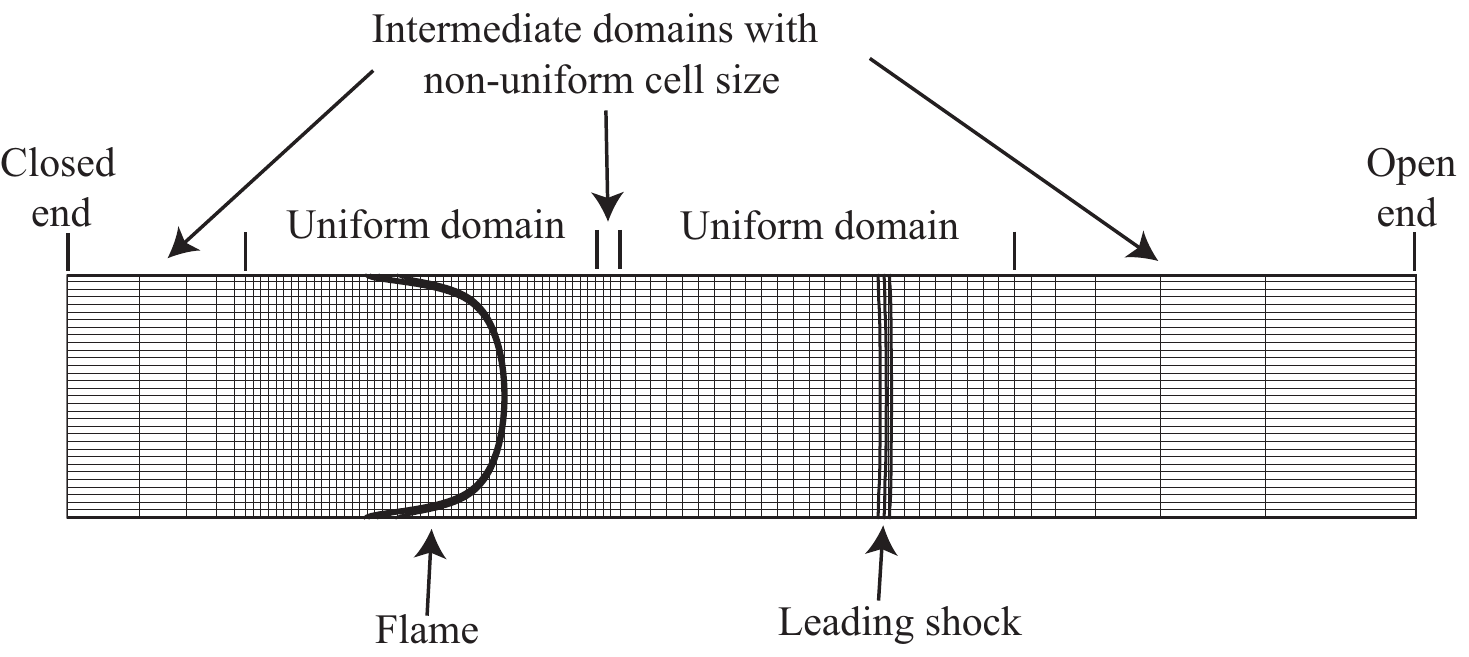}
\caption{The sketch of the grid with variable resolution used in
numerical simulations.} \label{fig-grid-sketch}
\end{figure}

It is emphasized that the result~(\ref{eq74}) does not reproduce the
asymptote~(\ref{eq62_IV}) as they are related to opposite limiting cases.
Consequently, it is expected that for a certain range of $Ma$ the complete (numerical) solution to
Eq.~(\ref{eq62_II}) lies in between the two values given by Eqs.~(\ref{eq62_IV})
and (\ref{eq74}).


\section{Numerical method, basic equations, boundary and initial conditions}
\label{Sec.num.method}

We perform numerical simulations of the hydrodynamic and
combustion equations including transport processes (thermal
conduction, diffusion, viscosity) and chemical kinetics in the
form of Arrhenius equation. Both 2D planar and axisymmetric
cylindrical flows are investigated. In the general tensor form the
governing equations are given by

\begin{eqnarray}
\label{eq77} {\frac{{\partial \rho}} {{\partial t}}}  +  {\frac{{1}}{{r^{\beta }}}}{\frac{{\partial }} {{\partial r}}}
\left( {r^{\beta} \rho u_{r}} \right)
 +   {\frac{{\partial }}
{{\partial z}}}\left( {\rho  u_{z}} \right) = 0,
\end{eqnarray}
\begin{eqnarray}
\label{eq78} 
\frac{\partial} {\partial t} \left( \rho u_{r}
\right) & + & \frac{1}{r^{\beta}} \frac{\partial }{\partial
 r}{\left[ {r^{\beta} \left( {\rho
u_{r}^{2} - \zeta _{r,r}}  \right)} \right]} + \nonumber \\
& + & {\frac{{\partial}} 
{{\partial z}}}\left( {\rho u_{r} u_{z} - \zeta _{r,z}} \right) +
{\frac{{\partial P}}{{\partial r}}} + \psi _{\beta}  = 0,  \nonumber \\
\end{eqnarray}
\begin{eqnarray}
\label{eq79} 
{\frac{{\partial}} {{\partial t}}}\left( {\rho u_{z}}
\right) & + & {\frac{{1}}{{r^{\beta}} }}{\frac{{\partial }}{{\partial
 r}}}{\left[ {r^{\beta} \left( {\rho u_{r}
u_{z} - \zeta _{r,z}}  \right)} \right]} + \nonumber \\
& + & {\frac{{\partial
}}{{\partial z}}}\left( {\rho u_{z}^{2} - \zeta _{z,z}}  \right) +
{\frac{{\partial P}}{{\partial z}}} = 0{\rm ,}
\end{eqnarray}
\begin{eqnarray}
\label{eq80} {\frac{{\partial \varepsilon}} {{\partial t}}} & + &
{\frac{{1}}{{r^{\beta }}}}{\frac{{\partial}} {{\partial
r}}}{\left[ {r^{\beta} \left( {\left( {\varepsilon + P}
\right)u_{r} - \zeta _{r,r} u_{r} - \zeta _{r,z} u_{z} + q_{r}}
\right)} \right]} + \nonumber \\
& + & {\frac{{\partial}} {{\partial
z}}}{\left[ {\left( {\varepsilon + P} \right)u_{z} - \zeta _{z,z}
 u_{z} - \zeta _{r,z}  u_{r} + q_{z}}
\right]} = 0,
\end{eqnarray}
\begin{eqnarray}
\label{eq81} \frac{\partial} {\partial t} \left( {\rho Y}
\right) & + & \frac{1}{r^{\gamma}} \frac{\partial} {\partial
 r} {\left[ {r^{\gamma} \left( {\rho
 u_{i} Y - \frac{\mu}{Sc} \frac{\partial Y}{\partial r}} \right)} \right]} + \nonumber \\
& + &
\frac{\partial} {\partial  z}\left( \rho u_{z} Y -
\frac{\mu}{Sc} \frac{\partial Y}{\partial
  z} \right) = \nonumber \\
& = & - \frac{\rho   Y}{\tau _{R}} \exp \left(  - E_{a} / \overline{R} T \right),
\end{eqnarray}
\noindent where $\beta = 0$ and 1 for 2D and axisymmetric
geometries, respectively,
\begin{equation}
\label{eq82}
\varepsilon = \rho \left( {QY + C_{V} T} \right) + {\frac{{\rho
}}{{2}}}\left( {u_{z}^{2} + u_{r}^{2}}  \right)
\end{equation}
\noindent is the total energy per unit volume, $Y$ the mass
fraction of the fuel, $Q$ the energy release from the reaction, and $C_{V} $ the heat capacity at constant volume. The energy
diffusion vector $q_{i} $ is given by
\begin{equation}
\label{eq83} q_{r} = - \mu \left( {{\frac{{C_{P}}} {{Pr}}
}{\frac{\partial T}{\partial r}} + \frac{Q}{Sc}{\frac{\partial Y}{\partial r}}} \right),
\end{equation}
\begin{equation}
\label{eq84} q_{z} = - \mu \left( {{\frac{{C_{P}}} {{Pr}}
}{\frac{{\partial T}}{{\partial   z}}} + {\frac{Q}{Sc}}{\frac{\partial Y}{\partial z}}} \right).
\end{equation}
In the 2D configuration ($\beta = 0)$ the stress tensor $\zeta _{i,j} $ takes the form
\begin{equation}
\label{eq85} \zeta _{i,j} = \mu \left( {{\frac{{\partial   u_{i}}}
{{\partial x_{j}}} } + {\frac{{\partial   u_{j}}} {{\partial
x_{i}}} } - {\frac{{2}}{{3}}}{\frac{{\partial   u_{k}}} {{\partial
x_{k} }}}\delta _{i,j}}  \right),
\end{equation}
\noindent  while in the
axisymmetric geometry ($\beta = 1)$ it reads
\begin{equation}
\label{eq86} \zeta _{r,r} = {\frac{{2\mu}} {{3}}}\left(
{2{\frac{{\partial u_{r}}} {{\partial   r}}} - {\frac{{\partial
u_{z} }}{{\partial   z}}} - {\frac{{u_{r}}} {{r}}}} \right),
\end{equation}
\begin{equation}
\label{eq87} \zeta _{z,z} = {\frac{{2\mu}} {{3}}}\left(
{2{\frac{{\partial u_{z}}} {{\partial   z}}} - {\frac{{\partial
u_{r} }}{{\partial   r}}} - {\frac{{u_{r}}} {{r}}}} \right),
\end{equation}
\begin{equation}
\label{eq88} \zeta _{r,z} = \mu \left( {{\frac{{\partial   u_{r}}}
{{\partial   z}}} + {\frac{{\partial   u_{z}}} {{\partial   r}}}}
\right).
\end{equation}
Finally, the last term in Eq. (\ref{eq78}) takes the form
\begin{equation}
\label{eq89} \psi _{\beta}  = {\frac{{2\mu}} {{3}}}\left(
{2{\frac{{u_{r}}} {{r}}} - {\frac{{\partial   u_{r}}} {{\partial
  r}}} - {\frac{{\partial   u_{z}}} {{\partial
  z}}}} \right)
\end{equation}
\noindent if $\beta = 1$, and $\psi _{\beta}  = 0$ if $\beta = 0$.
Here $\mu$ is the dynamic viscosity, and $Pr$ and $Sc$
the Prandtl and Schmidt numbers, respectively.

\begin{figure*}[tp]
{\includegraphics[width=0.30\textwidth\centering]{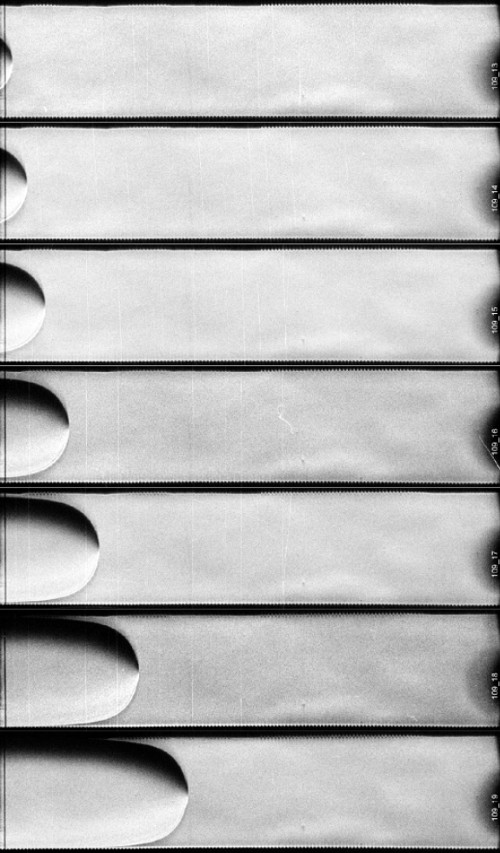}}
{\includegraphics[width=0.30\textwidth\centering]{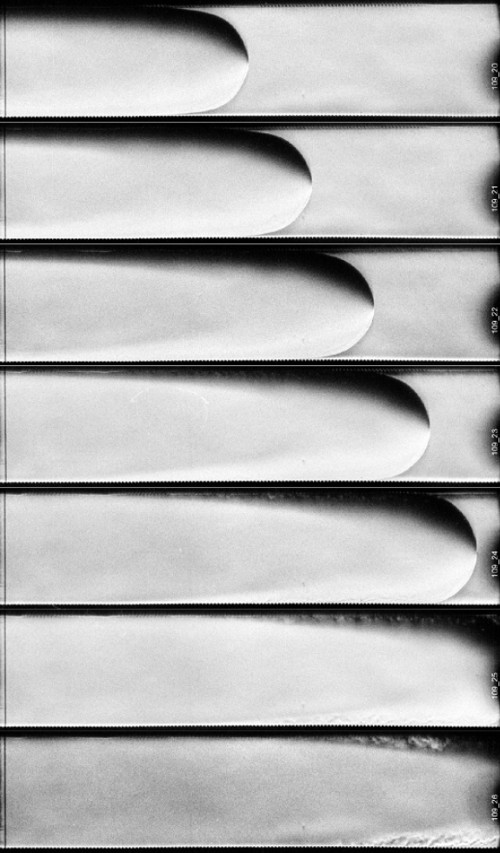}}
 \caption{Schlieren images of finger flame propagation for stoichiometric $H_2/O_2$ mixture at pressure 0.2 bar.
The shown images are evenly distributed in time with interval of 100 ms.
Note that the shift of image's left boundary is 10 mm to the right as compared to the position of the end wall and ignition point.}
 \label{fig-schlieren}
\end{figure*}

We take unity Lewis number $Le \equiv Sc/Pr = 1$, with $Pr  =
Sc = 0.75$; the dynamical viscosity is $\mu = 1.7\times 10^{ - 5}\,{\rm N}{\rm
s}{\rm /} {\rm m}^{{\rm 2}}$. The fuel-air mixture and burnt gas are
perfect gases with a constant molar mass $m = 2.9\times 10^{ - 2}{\rm kg}/ 
{\rm mol}$, with $C_{V} = 5\overline{R} / 2m$, $C_{P} =
7\overline{R} / 2m$, and the equation of state
\begin{equation}
\label{eq90}
P = \rho \overline{R} T / m,
\end{equation}
\noindent where $\overline{R} \approx 8.31 \,  \mathrm{J} / (\mathrm{
mol} \cdot \mathrm{K} )$ is the universal gas
constant. We consider a single-step irreversible reaction of the
first order with the temperature dependence of the reaction rate
given by the Arrhenius law with an activation energy $E_{a} $ and
the factor of time dimension $\tau _{R} $. In our simulations we
took $E_a / \overline{R} T_{u} = 32$ in order to have better resolution of
the reaction zone. The factor $\tau _{R} $ was adjusted to obtain
a particular value of the planar flame velocity $S_{L} $ by
solving the associated eigenvalue problem. The flame thickness is
defined as
\begin{equation}
\label{eq91} L_{f} \equiv {\frac{{\mu _{u}}} {{\Pr \rho _{u}
S_{L}}} },
\end{equation}
\noindent where $\rho _{u} = 1.16\,\rm{kg / m^{3}}$ is the
unburned mixture density. It is noted that $L_{f} $ is just a mathematical
parameter of length dimension related to the flame front, while
the real effective diffusion flame thickness is considerably larger
\cite{Poinsot.Veynante-2001,Akkerman.et.al-2006-2}.
We took initial temperature of the fuel mixture $T_{u, 0} = 300K$,
initial pressure $P_{u, 0} = 10^{5}\ Pa$, specific heat ratio
$\gamma = 1.4$, and $\Theta = 8$. We performed the
simulations for a rather wide range of initial Mach number $Ma = S_{L}
/ c_{S, 0} = 10^{ - 3} \sim 1.6\times 10^{-2}$, 
with the lower and upper values being relevant to hydrocarbon and hydrogen-oxygen flames, respectively
 \cite{Kuznetsov.et.al-2005}. We used the
tube diameter $2R = 150 L_{f} $  and channel width $2R = 100 L_{f} $ for axisymmetric 
and 2D simulations, respectively.

Similar to the theoretical analysis, we adopt slip and
adiabatic boundary conditions at the tube walls:
\begin{equation}
\label{eq92}
{\rm {\bf n}} \cdot {\rm {\bf u}} = 0,
\quad
{\rm {\bf n}} \cdot \nabla T = 0,
\end{equation}
\noindent where ${\rm {\bf n}}$ is the unit normal vector at the
walls. At the open face end of the tube/channel non-reflecting boundary
conditions are applied. As initial conditions, we used a
semi-circular flame ``ignited'' at the channel
axis at the closed end of the tube, with its structure given by
the analytical solution of Zel'dovich and Frank-Kamenetskii
\cite{Zeldovich.et.al.-1985, Law-2006}
\begin{eqnarray}
\label{eq93-1}
& & T = T_{u} + (T_{b} - T_{u} )\exp \left( { - \sqrt {x^2 + z^2} / L_{f}}
\right), \nonumber \\ 
& & \textrm{if}\ z^{2} + x^{2} < r_{f}^{2}
\end{eqnarray}
\begin{equation}
\label{eq93-2} T = \Theta T_{u} ,\ \textrm{if}\ z^{2} + x^{2} >
r_{f}^{2}
\end{equation}
\begin{eqnarray}
\label{eq93-3} 
& & Y = (T_{b} - T) / (T_{b} - T_{f} ), \quad P = P_{u}, \nonumber \\
& & u_{x} = 0, \quad u_{z} = 0.
\end{eqnarray}
Here $r_{f}$ is the radius of the initial flame ball at the closed end
of the tube. The finite initial radius of the flame ball is
equivalent to a time shift, which requires proper adjustments when
comparing the theory and numerical simulations.

The simulations used a 2D hydrodynamic Navier-Stokes code
adapted for parallel computations \cite{Wollblad.et.al-2006}. The numerical scheme is second
order accurate in time and fourth order accurate in space for convective
terms, and second order in space for diffusive terms. The code is
robust and accurate; it was successfully used in aero-acoustic
applications. 2D and
axisymmetric simulations were conducted. We used  mesh with variable resolution
in order to take into account the growing distances between the
tube end, the accelerating flame and the pressure wave, and to
resolve both chemical and hydrodynamic spatial scales. Typical
computation time for one simulation required up to $10^{4}$
CPU-hours, hence implying the need for extensive  parallel
calculations.

A rectangular grid with the grid walls parallel to the
coordinate axes was used. The sketch of the calculation mesh used in
simulations of flame acceleration from the closed tube end is
shown in Fig. \ref{fig-grid-sketch}. To perform all the
calculations in a reasonable time, we made the grid spacing
non-uniform along the $z$-axis with the zones of fine grid around
the flame and leading shock fronts. For  majority of the simulation runs, the grid size in the $z$-direction was
0.25$L_{f} $ and 0.5$L_{f}$ in the domains of the flame and leading pressure wave, respectively, which
allowed resolution of the flame and
waves. Outside the region of fine grid the mesh size increased
gradually with 2\% change in size between the neighboring cells.
In order to keep the flame and pressure waves in the zone of fine
grid we implemented the periodical mesh reconstruction during the
simulation run \cite{Valiev.et.al-2008}.
Third-order splines were used for the re-interpolation of the
flow variables during periodic grid reconstruction to preserve the
second order accuracy of the numerical scheme.


\section{Results and discussion}

\subsection{Experimental results}
\label{Sec.experiment}

\begin{figure}
\includegraphics[width=0.46\textwidth\centering]{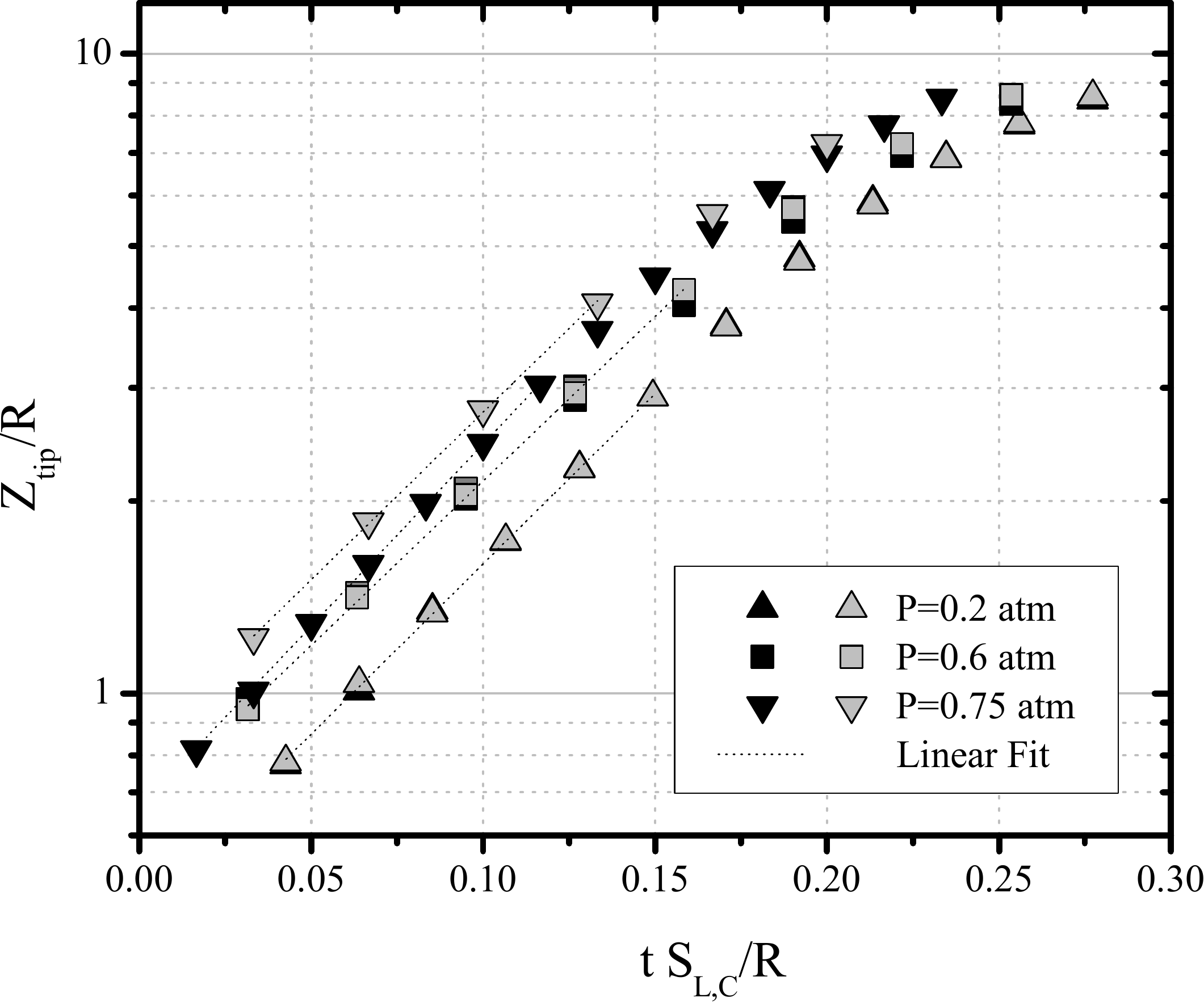}
 \caption{Scaled flame tip position versus scaled time for different pressures:
  $P=0.2\ \textrm{bar}$ (triangles),  $P=0.6\ \textrm{bar}$ (squares),  $P=0.75\
\textrm{bar}$ (upside down triangles).
 Different color for each pressure correspond to different experimental runs.}
 \label{fig-tip-pos-experiment}
\end{figure}

Experiments were performed in a channel with rectangular
cross-section (50 $\times $ 50 mm), 6.05 m long with 24
transparent ports for photo-gauges. A high-speed schlieren system
with stroboscopic pulse generator and high speed camera, germanium
photodiodes and piezoelectric transducers were used to record the
flame evolution. The experimental facilities, including ignition conditions, are described in
detail in Ref. \cite{Kuznetsov.et.al-2005, Kuznetsov.et.al-2010} and the references given therein. 

\begin{table*}[tp]
\begin{tabular}{|c|c|c|c|c|c|c|c|c|c|}
\hline \ $P$, bar \ & \ $\Theta $ \ & \ $L_T$, mm \ &\ $C $\ & \ $S_L$, m/s \ & \ $S_{L,C}$, m/s \
&\ $Ma$ \ & \ $Ma_C$ \ &\ $\sigma_1 $ &\ $\sigma_{1,C} $\
 \\
\hline
0.20 &\ \ 8.00 \ \ &\ \ 2.0\ \ &\ \ 0.778\ \ & 6.855 & 5.33 &\ \ 0.0128\ \ &\ \ 0.0100\ \ &\ \ 4.18\ \ &\ \ 5.373\ \ \\
0.60 &\ \ 8.260\ \ &\ \ 0.5\ \ &\ \ 0.944\ \ & 8.395 &7.928 &\ \ 0.0157\ \ &\ \ 0.0149\ \ &\ \ 4.83\ \ &\ \ 5.138\ \ \\
0.75 &\ \ 8.317\ \ &\ \ 0.379\ \ &\ \ 0.958\ \ & 8.700 &8.335 &\ \ 0.0162\ \ &\ \ 0.0157\ \ &\ \ 4.90\ \ &\ \ 5.110\ \ \\
0.75 &\ \ 8.317\ \ &\ \ 0.379\ \ &\ \ 0.958\ \ & 8.700 &8.335 &\ \ 0.0162\ \ &\ \ 0.0157\ \ &\ \ 5.43\ \ &\ \ 5.660\ \ \\
\hline
\end{tabular}
\caption{Experiment parameters and $\sigma_{1,C}$ obtained by fitting
of the results of Fig. \ref{fig-tip-pos-experiment} for different
pressures.} \label{table-experiment-comparison}
\end{table*}

\begin{figure}
\includegraphics[width=0.45\textwidth\centering]{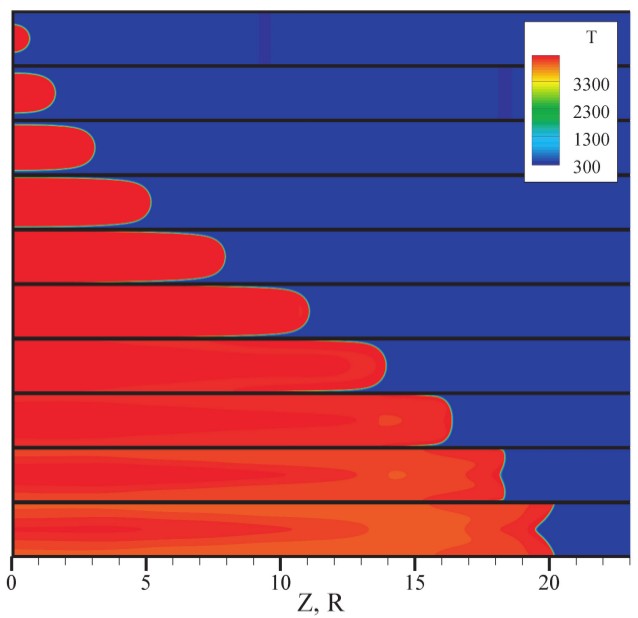}
\caption{Temperature field evolution and ``tulip'' formation for
planar geometry, $\Theta= 14$, and Mach number $Ma=0.005$ at
different time instants. Time instants are equally spaced in the
range of $(0.04022 \div 0.42161) S_L t/R$. }
 \label{fig-2d-shapes}
\end{figure}

The experiments were conducted with the fast-burning,
stoichiometric hydrogen--oxygen mixture at initial pressures of $0.2 \sim 0.75$
bar. The mixture was prepared by precise partial pressure method with deviation less than 
$\pm 0.2\% $ with respect to the $H_2$ fraction.
By changing the initial pressure we vary the value of $S_L$ and the  initial flame
Mach number. The initial ambient gas mixture temperature was $T_0=293 K$; and the corresponding sound
speed is 531 m/s. The initial density ratio is in the range $\Theta=8.0 \sim 8.317$,
depending on the initial pressure.
It is noted that for such highly 
reactive fuel mixtures the influence of the ignition source on the initial flame dynamics is negligible, as compared 
with the influence of mixture reactivity, expansion ratio and tube diameter. 
The ignition energy of 2-10 mJ is about $10^3$ times lower than the released 
combustion energy at the initial stage of finger flame propagation (with the diameter of flame ball smaller than 1 cm).

\begin{figure*}
\includegraphics[width=0.5\textwidth\centering]{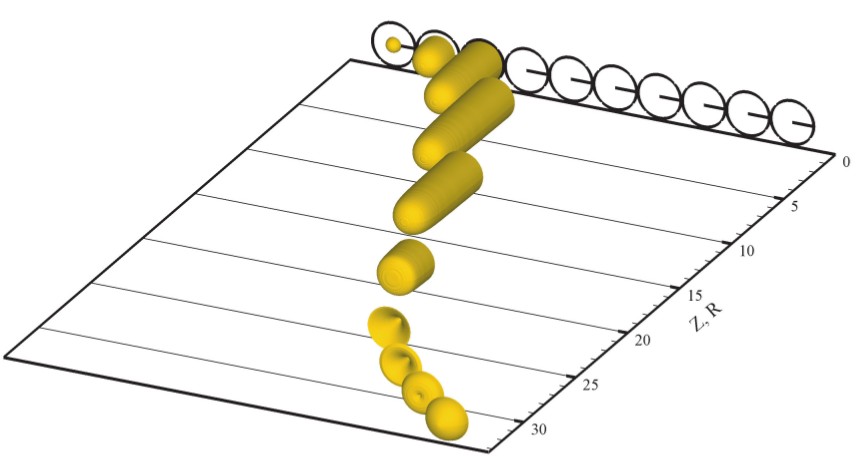}
\caption{Flame shape evolution and ``tulip'' formation for
axisymmetric geometry, $\Theta= 14$, and Mach number $Ma=0.005$ at
different time instants. Time instants are equally spaced in the
range of $(0.02692
 \div 0.26576 ) S_L t/R$. Isosurfaces are shown for $T=1400 K$.}
 \label{fig-3d-shapes}
\end{figure*}

Figure~\ref{fig-schlieren} shows the schlieren images of ``finger''
flame propagation at a pressure of 0.2 bar.
It is seen that the flame tip accelerates
exponentially at the initial stage of flame propagation, in agreement with previous
experiments \cite{Clanet.Searby-1996} and theory \cite{Bychkov.et.al-2007}.
For the stoichiometric $H_2/O_2$ fuel mixture used in the present experiment the influences 
of diffusional-thermal
cellular and pulsating instabilities \cite{Law-2006} are ruled out. 
The influence of hydrodynamic (Darrieus-Landau) instability  is
negligible as well due to the strong flame curvature  observed in the experiment, 
leading to the Zeldovich-type stabilization of
the flame front perturbations  \cite{Liberman.et.al-2003}.
Moreover, even if the Darrieus-Landau instability is developed, 
its scaled exponential acceleration rate would be about unity \cite{Zeldovich.et.al.-1985,Law-2006}, 
which is up to an order of magnitude smaller than the finger 
flame acceleration rate of Eqs. (\ref{eq19-1}), (\ref{eq45-1}), (\ref{2.3.eq23-1}), (\ref{eq69}), 
thereby diminishing any possible role of the DL instability as compared 
to the finger-flame acceleration effects.     
Consequently, in the present experiment the flame front acceleration could be attributed purely to the finger-flame mechanism
of exponential acceleration \cite{Clanet.Searby-1996,Bychkov.et.al-2007}.

\begin{figure*}
\subfigure[]{\includegraphics[width=0.43\textwidth\centering]{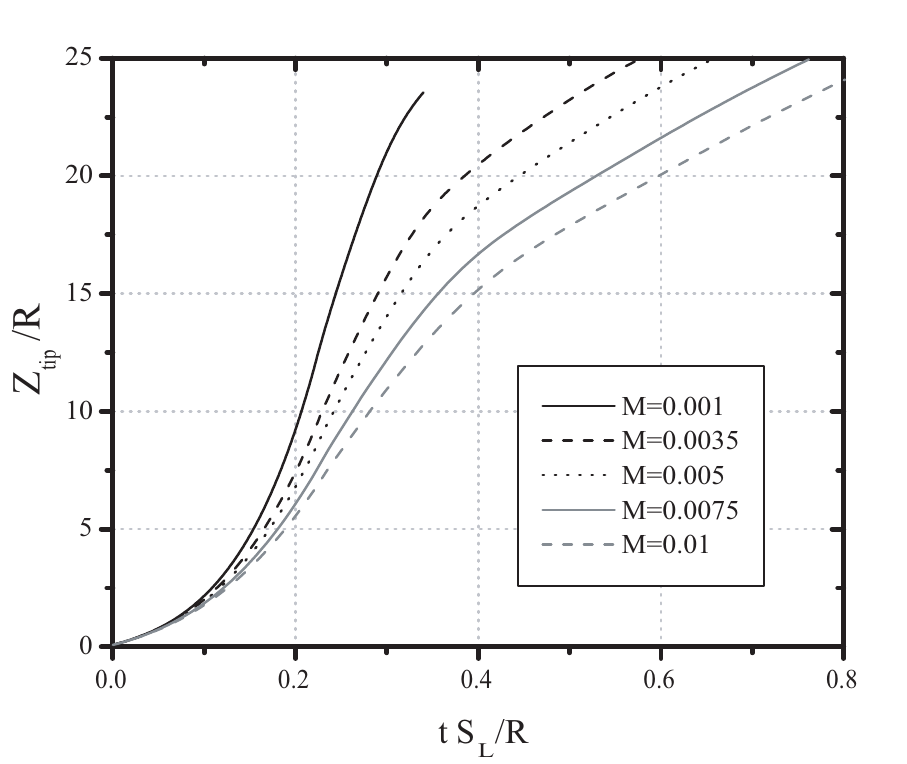}}
\subfigure[]{\includegraphics[width=0.43\textwidth\centering]{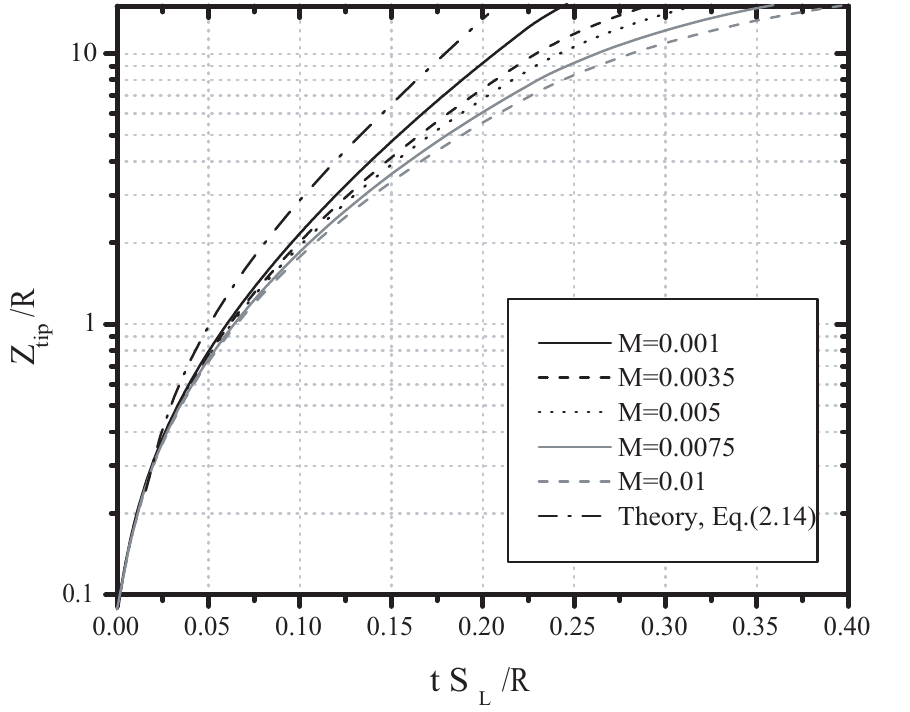}}
 \caption{Scaled flame tip position versus time for planar geometry, $\Theta=
14$, and Mach numbers $Ma=0.001,\ 0.0035,\ 0.005,\ 0.0075,\ 0.01$: (a) linear
scale, (b) logarithmic scale.}
 \label{fig-ztip.planar.theta14}
\end{figure*}

The experimentally obtained evolution of the scaled flame tip position for various pressures
 is shown in Fig.~\ref{fig-tip-pos-experiment}. The definition of the laminar burned velocity $S_{L,C}$ used in the
present scaling is described further. When the
flame skirt approaches the wall, the flame acceleration weakens,
and the flame shape undergoes transition from a convex ``finger'' to a
concave ``tulip'' shape. The instant when the flame skirt approaches the channel
wall is clearly correlated with the instant at which the exponential flame tip
acceleration terminates, which can be seen at Fig.~\ref{fig-tip-pos-experiment}.

\begin{figure}
\includegraphics[width=0.43\textwidth\centering]{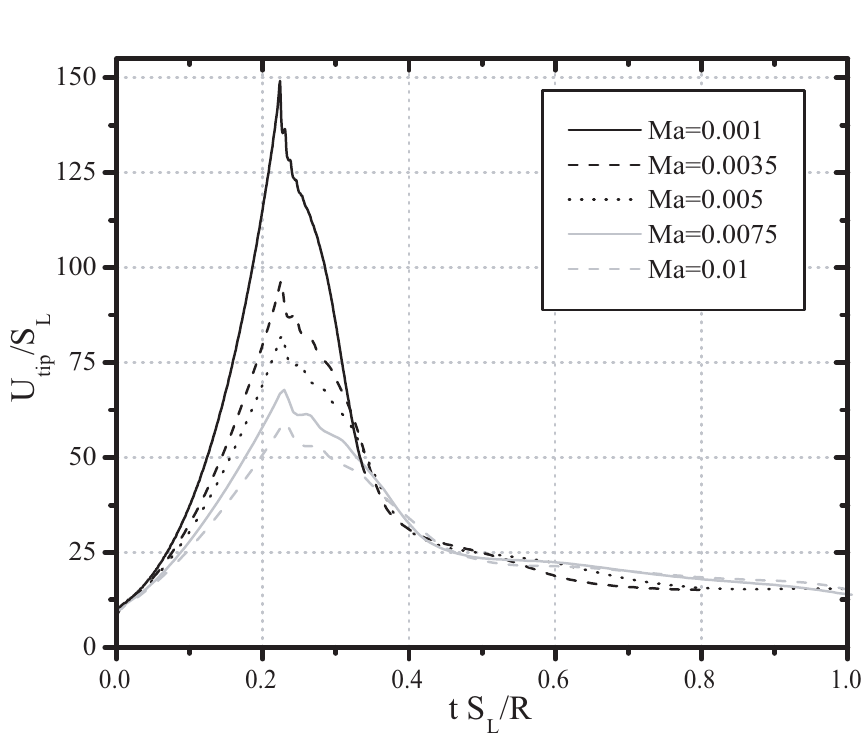}
 \caption{Scaled velocity of flame tip versus time for planar geometry, $\Theta=
14$, and Mach numbers $Ma=0.001,\ 0.0035,\ 0.005,\ 0.0075,\ 0.01$.}
 \label{fig-utip.planar.theta14}
\end{figure}

Physical parameters of the experiments, as well as the scaled acceleration rates
obtained from fitting of experimental results of
Fig.~\ref{fig-tip-pos-experiment} for different pressures, and the
initial Mach numbers are given in Table~\ref{table-experiment-comparison}.
The unstretched laminar flame
velocity $S_L$ and the thermal flame thickness $L_{T}$ are obtained from the numerical simulation of one-dimensional premixed flame structure
employing PREMIX code of the CHEMKIN family \cite{Kee-2000} with the use of updated
chemical kinetics mechanism for hydrogen oxidation \cite{Burke-2011}.
The thermal flame thickness $L_{T}$ is conventionally defined as \cite{Law-2006} $(T_b-T_u)/\max(|\partial T/\partial x|)$, where $T_b$ and $T_u$ are the
temperatures of burnt and unburnt gases, respectively, and $\max(|\partial T/\partial x|)$ is the maximum of the
temperature gradient. It is noted that in our case of low-pressure stoichiometric hydrogen-oxygen finger flames a noticeable
pure curvature effect \cite{Law-2006} is observed, therefore we introduce relevant modifications to
$S_L$ and $Ma$, denoted as $S_{L,C}$ and $Ma_C$, in such a mannner that the correction parameter $C$ is defined as $C=S_{L,C}/S_{L}$, so that
the estimation for the modified
growth rate is given by $\sigma_{1,C}=\sigma_1/C$.
It is seen that, the lower is the pressure, the larger is the thermal flame thickness, 
and consequently the more pronounced is the pure curvature effect.

\begin{figure}
\includegraphics[width=0.45\textwidth\centering]{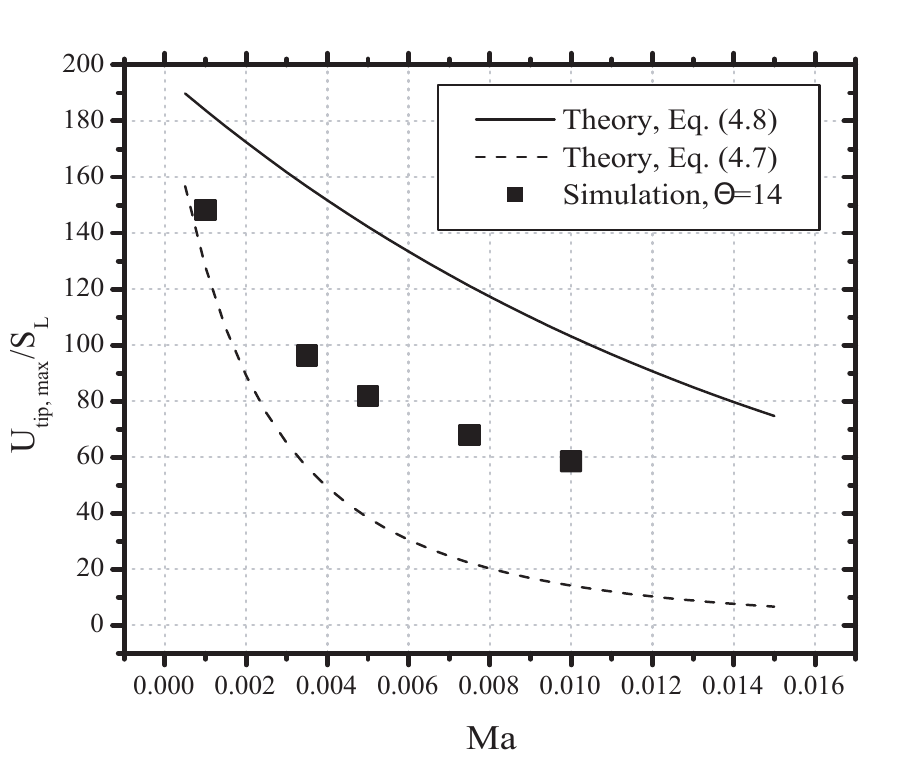}
 \caption{Scaled maximum flame tip velocity versus Mach number obtained in simulations for planar
geometry, $\Theta= 14$, is compared to theoretical estimates. Solid line -
Eq. (\ref{eq.utipmax.2d.begin}), dashed line - Eq. (\ref{eq.utipmax.2d})}
 \label{fig-utipmax.planar.theta14}
\end{figure}

The pure curvature correction parameter $C$ and the modified growth rate $\sigma_{1,C}$ are estimated as follows. Based on experimental
observation, we assume, realistically, that the curvature of the flame tip remains almost constant
during the time interval $\tau_{sph}<\tau<\tau_{wall}$, with $\tau_{sph}$ and
 $\tau_{wall}$ given by Eqs. (\ref{eq14}) and (\ref{eq16}), respectively.
Within this time the flame tip radius can be approximated as
\begin{equation}
 R_{tip}= R \eta_{f,sph} = 0.63 \, R \Theta / \left(\Theta-1\right) \approx 0.72 R,
\label{eq.exper-tip-radius}
\end{equation}
\noindent where $R$ is the channel
half-width. Consequently, with the curvature term in the form $\nabla\cdot \mathbf{n}=2/R_{tip}$,
the curvature-modified laminar burning velocity at the flame tip can be estimated as \cite{Law-2006}
\begin{equation}
 S_{L,C}=C S_L = S_L \left(1- L_T \nabla\cdot \mathbf{n}\right) = S_L \left(1- \frac{2 L_T}{R_{tip}}\right).
\label{eq.SL_corr}
\end{equation}
\noindent  As Table \ref{table-experiment-comparison} shows, 
modified values of the growth rate $\sigma_{1,C}$ decrease with increasing $Ma$,
while the uncorrected growth rate $\sigma_1$ increases instead. This demonstrates a non-negligible effect of pure curvature
for low-pressure $H_2/O_2$ finger
flames, disregarding which could reverse the main trend. For this reason,
an accurate determination of the laminar burning velocity
is crucial for the analysis of experimental results. This is particularly relevant for the extraction of the growth
rate $\sigma_1$ for various $Ma$, since even a 15-20\%
difference in the laminar burning velocity could lead to a completely erroneous conclusion.

\subsection{Numerical results and discussion}
\label{Sec.num.results}

We performed numerical simulation of the flame acceleration in
tubes with smooth slip adiabatic walls at different flow
parameters. Particularly, we employed planar and axisymmetric
geometries, and a range
of initial flame propagation Mach numbers $Ma = 10^{ - 3} \sim 1.6
\times 10^{ - 2}$. We also used two values of the thermal expansion
ratio $\Theta=8, \ 14$.

\begin{figure*}
\subfigure[]{\includegraphics[width=0.43\textwidth\centering]{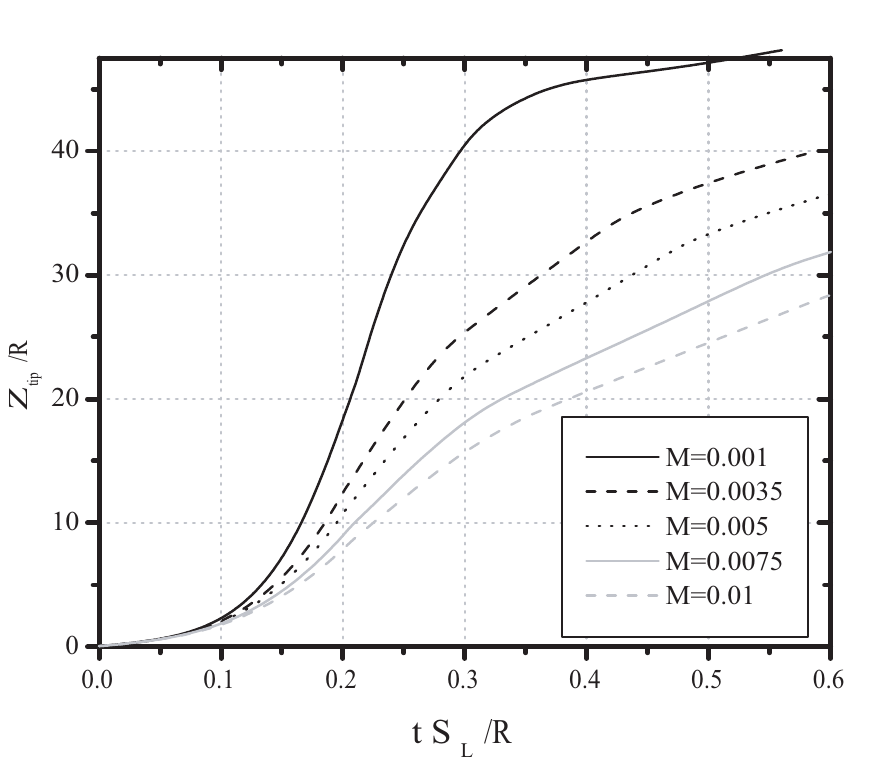}}
\subfigure[]{\includegraphics[width=0.43\textwidth\centering]{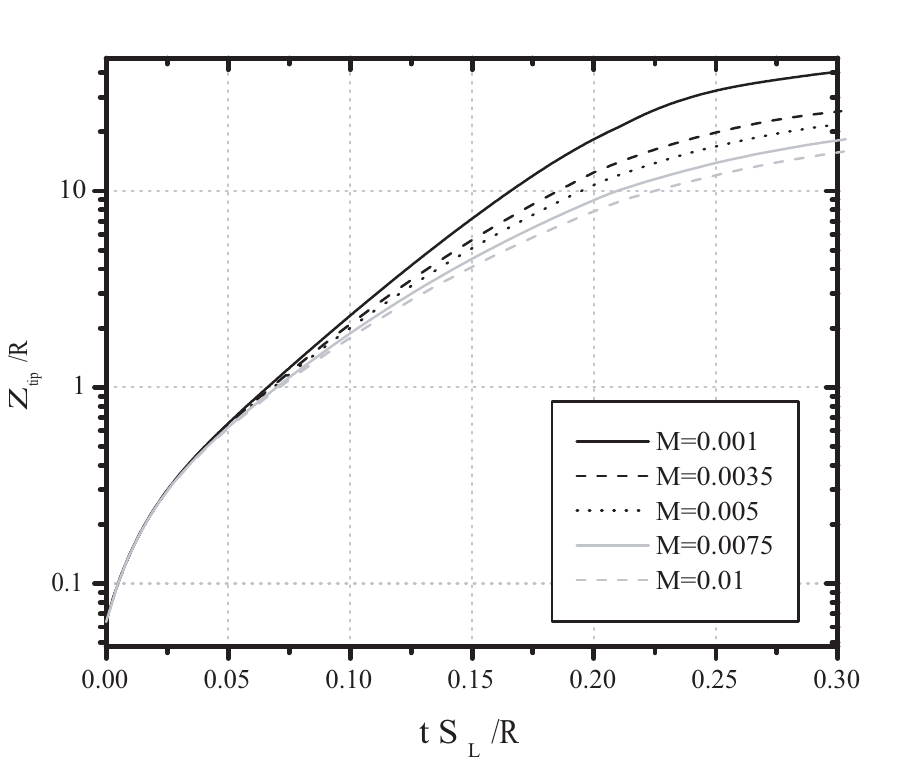}}
 \caption{Scaled flame tip position versus time for axisymmetric geometry,
$\Theta= 14$, and Mach numbers $Ma=0.001,\ 0.0035,\ 0.005,\ 0.0075,\ 0.01$: (a)
linear scale, (b) logarithmic scale.}
 \label{fig-ztip.axisym.theta14}
\end{figure*}

 We illustrate evolution of the temperature
field for the planar geometry in Fig. \ref{fig-2d-shapes} for  $\Theta= 14$ and the initial Mach number
$Ma=0.005$. The simulation plots also show the instant
of transition from a convex ``finger'' to a concave ``tulip''
flame shape. It is seen that, similar to the experimental Figure \ref{fig-schlieren}, the flame tip curvature does not
change significantly from the instant
of the transition to the finger configuration and until the flame skirt touches the wall, thus justifying the assumption of almost constant tip curvature
made in Sec. \ref{Sec.experiment}. It should be noted that  formation
of finger-shaped laminar flame fronts in planar geometry due to essentially different Schelkin mechanism has been also obtained in simulations
of premixed flames in channels with non-slip walls \cite{Kagan-2003, Valiev.et.al-2008, Valiev.et.al-2009}. 
Furthermore similar shapes of the finger front 
were observed  within the context of
electrochemical doping in organic semiconductors \cite{Bychkov.et.al-2011,Bychkov.et.al-2012}, with the electric field playing  conceptually the same role as the field of gas velocity in the present combustion problem.
Another interesting physical example of front acceleration and DDT has been encountered recently in the studies of spin-avalanches 
in crystals of nanomagnets \cite{Decelle.et.al-2009,Modestov.et.al-2011}.

\begin{figure}
\includegraphics[width=0.45\textwidth\centering]{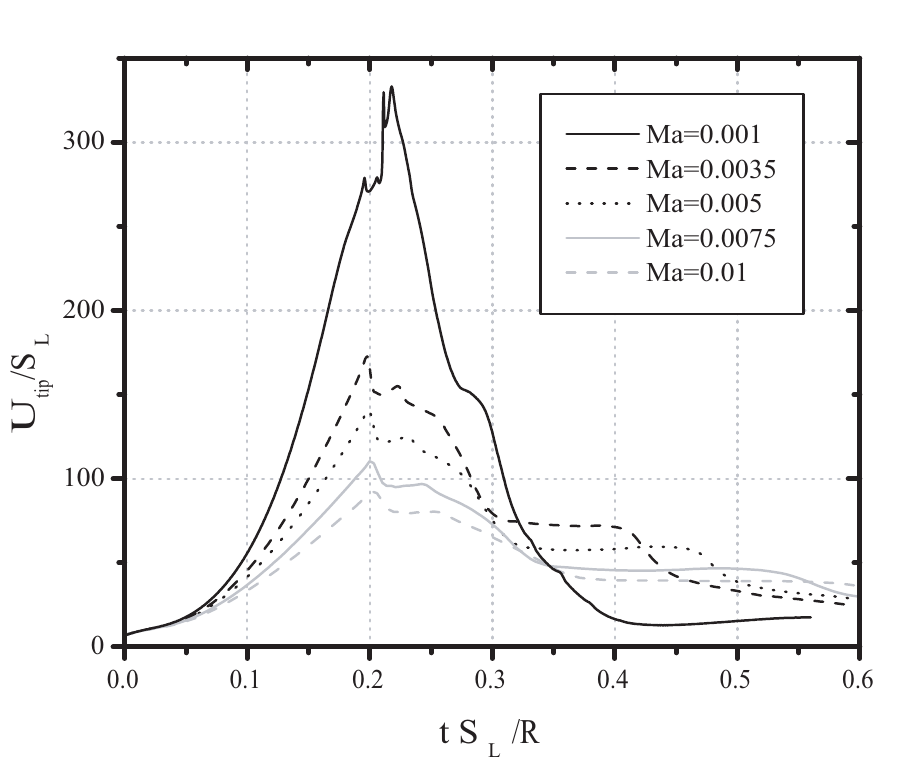}
 \caption{Scaled velocity of flame tip versus time for axisymmetric geometry,
$\Theta= 14$, and Mach numbers $Ma=0.001,\ 0.0035,\ 0.005,\ 0.0075,\ 0.01$.}
 \label{fig-utip.axisym.theta14}
\end{figure}

  An axisymmetric counterpart of Fig. \ref{fig-2d-shapes} is shown in Fig. \ref{fig-3d-shapes} with all parameters being the same except for
the  geometry. Isosurfaces are shown for
$T=1400 K$. The axisymmetic simulation also demonstrates the transition from a
convex, ``finger'' flame front, to a concave ``tulip'' flame, accompanied by a significant reduction of the flame surface area and propagation velocity.
Similar to Figs. \ref{fig-schlieren} and \ref{fig-2d-shapes}, the flame tip curvature
remains almost constant in a range of scaled times, $\tau_{sph}<\tau <\tau_{wall}$, corresponding to the later stage of finger flame acceleration.

In Figure \ref{fig-ztip.planar.theta14} we present   the scaled flame tip position
versus time for the planar geometry with $\Theta= 14$ and $Ma=0.001 \sim 0.01$ in both linear and
logarithmic scales. We see that the ``exponential'' nature of acceleration in the planar case is not very pronounced in the
early stages, due to the relatively high value of $\tau_{sph} \approx 1/(\Theta -1)$
 compared to its axisymmetric counterpart $\tau_{sph} \approx 1/(2 \sqrt{\Theta(\Theta-1)})$ \cite{Bychkov.et.al-2007};
the latter being approximately twice smaller for high values of $\Theta$ than in the planar case.
Figure \ref{fig-utip.planar.theta14} shows the scaled velocity
evolution for $\Theta=14$ and the planar geometry for a set of Mach numbers in the range $Ma=0.001 \sim
0.01$. Here we see a significant dependence of the maximum flame tip velocity
on the initial Mach number. At the same time,  Fig.
\ref{fig-utip.planar.theta14} shows that the scaled time of attaining
the velocity maximum is almost independent of the Mach number, which can be  demonstrated analytically, as follows.
If we write the first-order correction to $\xi_{tip}$ in the
form $\xi_{tip}=\xi_{tip,0}+Ma\xi_{tip,1}$, where $\xi_{tip,0}$ is
given by Eq.~(\ref{eq19}), then Eq.~(\ref{eq35}) becomes
\begin{eqnarray}
\label{eq35.mod} \vartheta & = & \Theta - Ma(\gamma - 1)(\Theta -
1)^{2}\left( \xi_{tip,0}+Ma\xi_{tip,1} + 1 \right) \approx \nonumber \\ & \approx &
\Theta - Ma(\gamma - 1)(\Theta - 1)\left[ \Theta
\exp{((\Theta-1)\tau)}-1 \right]. \nonumber \\ 
\end{eqnarray}
Similarly, replacing $\tau$ in Eq. (\ref{eq35.mod}) by $\tau_{wall}$ in the form
$\tau_{wall}=\tau_{wall,0}+Ma\tau_{wall,1}$, we find
\begin{equation}
\label{eq35.wall.mod} \vartheta_{wall} \approx  \Theta - Ma(\gamma
-1)(\Theta - 1)(\Theta^2-1).
\end{equation}
Substituting Eq.~(\ref{eq35.wall.mod}) into Eq.~(\ref{eq16}), we
obtain the estimation of $\tau_{wall}$ in the compressible case for the
planar geometry:
\begin{eqnarray}
\label{eq.tau.wall.comp}
\tau_{wall} & = & \frac{\ln\vartheta}{\vartheta-1} \approx \nonumber \\
& \approx &
\frac{\ln\Theta}{\Theta-1}\left[ 1+ Ma(\gamma -1)(\Theta^2 -
1)\left(1-\frac{\Theta-1}{\Theta\ln\Theta} \right)\right]. \nonumber \\
\end{eqnarray}

\begin{figure}
\includegraphics[width=0.45\textwidth\centering]{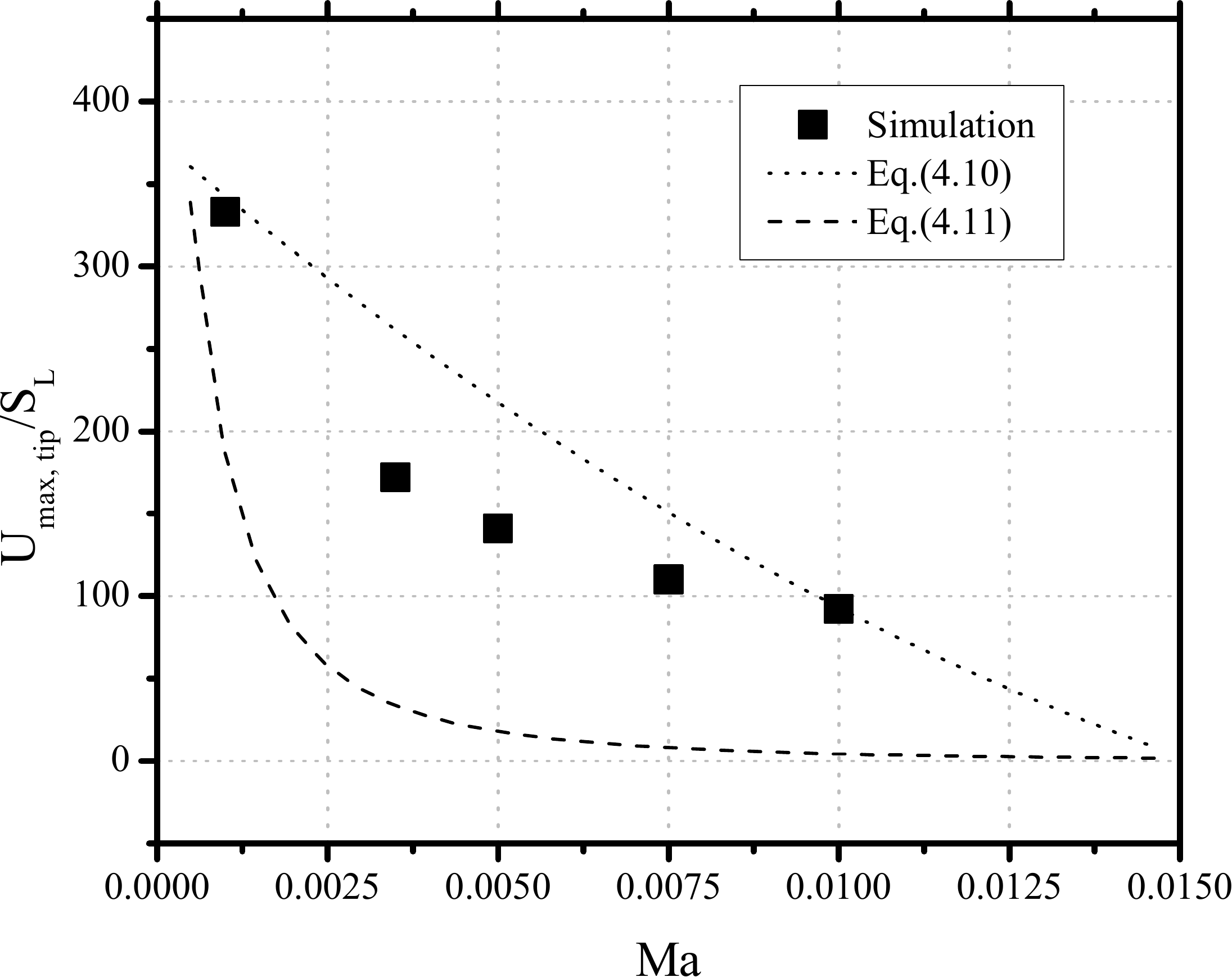}
 \caption{Scaled maximum flame tip velocity versus Mach number obtained in simulations for axisymmetric
geometry, $\Theta= 14$, compared to theoretical estrimates of Eqs.~(\ref{eq.utipmax.axi.begin}) and (\ref{eq.utipmax.axi.end}).}
 \label{fig-utipmax.axisym.theta14}
\end{figure}

For typical $\Theta \approx 5 \sim 10$ and $\gamma \approx 1.4$, the last term in Eq. (\ref{eq.tau.wall.comp}) can be approximated
as $\sim 0.2 Ma \Theta^2$. Consequently,
for small Mach numbers, $Ma \ll 5/\Theta^2 \sim 0.1$, the quantity $\tau_{wall}=\ln\vartheta/(\vartheta-1)$
only slightly depends on $Ma$, which is substantiated by the
simulation results of Fig.~\ref{fig-utip.planar.theta14}.
The weak dependence of $\tau_{wall}$ on $Ma$ is convenient for
evaluating  the total time of the finger flame acceleration. We see from the simulation results of Figure \ref{fig-utip.planar.theta14} that
the instant of the maximum flame tip velocity $\tau_{wall}$ only slightly
increases with increasing $Ma$, remaining in the range $\tau_{wall} = 2.2 \sim 2.3$.
Using the simplified analytical expression for $\tau_{wall}$ we can also
estimate the maximum flame tip velocity from Eq.~(\ref{eq44}) and
Eq.~(\ref{eq46}).

The peculiar feature of concurrent maximums of flame tip velocity for various initial Mach numbers can act as a useful criterion for the validity
of $S_L$ and $Ma$ determined in the experiments. While in the numerical simulation we set $Ma$ as a parameter and obtain the concurrence of the
maximum tip velocities mentioned above implicitly, analysis of the experimental data can encounter significant difficulties due to
the uncertainty of $S_L$, as discussed in
\ref{Sec.experiment}. Since $S_L$ is used in scaling time and velocity, even a small inaccuracy in its determination could lead to considerable
shifting of the maximums of the tip velocity relative to each other. Thus, if the experimental conditions imply $R \approx R_{tip}\gg L_f$, i.e.
pure curvature effect on flame tip velocity
is insignificant at the later stages of the flame acceleration, the concurrence of the scaled flame tip velocity peaks could serve as an indication of correctly
determined $S_L$ values for different $Ma$.

It is further noted that, similar to the experiments, in the present numerical simulations the initial flame velocity of a
hemispherical flame front is considerably affected
by the pure curvature effect, which can be seen from Fig. \ref{fig-utip.planar.theta14} for the flame tip velocity taken at $t S_L/R=0$.
The initial flame radius in the numerical simulation is equal to $R_{init}=4.0 L_f$; thus, similar to Eq. \ref{eq.SL_corr}, we can
estimate the correction to the initial flame velocity $S_{L,\,init}$ in the planar case, with the curvature term $\nabla \cdot \mathbf{n}=1/R_{init}$
 in the form:
\begin{equation}
\label{eq.sl.init}
\frac{S_{L,\,init}}{S_{L}}=1-\frac{L_f}{R_{init}}=\frac{3}{4},
\end{equation}
\noindent For $\Theta=14$, the initial tip velocity in the laboratory reference
frame is $U_{tip}/S_L=\Theta S_{L,\,init}/S_{L}=0.75\Theta =10.5$, which is close to that observed in Fig. \ref{fig-utip.planar.theta14}. However,
the effect of pure curvature does not affect the present numerical simulations considerably, since
we have taken relatively large channel widths, corresponding to $R=50 \, L_f$ in the planar case, as specified in Sec. \ref{Sec.num.method}, which
renders pure curvature effects
to be negligible for all simulation runs.

Figure \ref{fig-utipmax.planar.theta14} shows the scaled maximum
flame tip velocity versus the Mach number for the planar geometry and
$\Theta= 14$.
Analytical estimates for the maximum flame tip velocity shown in Fig. \ref{fig-utipmax.planar.theta14} are calculated as follows.
The first theoretical estimate for the maximum flame tip velocity
accounts for all nonlinear terms of Eq. (\ref{eq44}) and employs the exact solution, Eq. (\ref{eq48}), as
\begin{eqnarray}
\label{eq.utipmax.2d}
\frac{U_{tip,\,max}}{S_L} & = & - Ma\gamma \left( {\Theta - 1} \right)^{2}\xi _{tip,\, wall}^{2} + \nonumber \\
& + & \sigma _{1,pl} \xi_{tip,\, wall}
+ \Theta_{1},
 \end{eqnarray}
\noindent where $\xi_{tip,\, wall}$ is the flame tip position at time $\tau_{wall}$ calculated from Eq. (\ref{eq48}), with $\tau=\tau_{wall}$.
The second estimate  is somewhat simplified accounting
 for the linear term only, derived from Eq. (\ref{eq46}) as 
\begin{equation}
\label{eq.utipmax.2d.begin}
\frac{U_{tip,\,max}}{S_L}  = \sigma _{1,pl} \xi_{tip,\, wall}+ \Theta_{1},
\end{equation}
\noindent with $\xi_{tip,\, wall} =  (\Theta_1/\sigma_{1,pl})\left[ \exp{(\sigma_{1,pl} \tau_{wall})} -1 \right]$.
It is seen from Fig. \ref{fig-utipmax.planar.theta14} that  predictions of the complete analytical solution,
Eq. (\ref{eq.utipmax.2d}), and the solution
accounting  for the
linear term only, Eq. (\ref{eq.utipmax.2d.begin}), differ significantly, with the results of numerical simulation lying closer
to the exact solution of Eq. (\ref{eq.utipmax.2d}).
 Most importantly, we see that both the simulation and theoretical results
show significant reduction of the scaled maximum flame tip velocity
with increasing initial Mach number, despite  the fact that the non-scaled
 maximum tip velocity could be still increasing.
Figure \ref{fig-utipmax.planar.theta14} shows that the influence of gas compressibility is noticeable for the estimation of the maximum flame
velocity for finger-type flame acceleration, while previous theoretical studies \cite{Clanet.Searby-1996, Bychkov.et.al-2007} did not account for
the significant reduction of the maximum scaled tip velocity for high initial Mach numbers. This is because they were
conducted for slow methane-air flames, for which the effect of compressibility did not 
manifest itself.

\begin{figure}
\includegraphics[width=0.47\textwidth\centering]{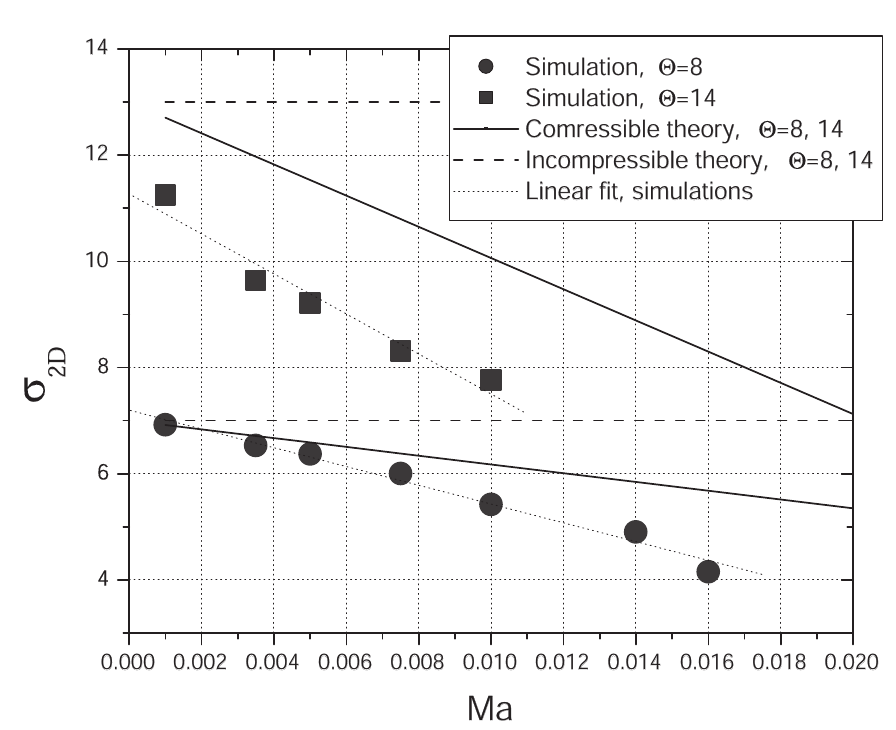}
 \caption{Scaled acceleration rate versus Mach number for planar geometry,
$\Theta=8, \
 14$. Values of $\sigma_{1}$ obtained from numerical simulations
 are shown by circles for $\Theta=8$ and by squares for
 $\Theta=14$.
 Solid lines correspond to the theoretical dependencies give by Eq.
(\ref{eq45-1}),
 dashed line - to incompressible theory, see Eqs. (\ref{eq18})-(\ref{eq19-1}),
 dotted lines - to linear fit of simulation data.}
 \label{fig-sigma-Ma-2D}
\end{figure}

We next investigate acceleration of the finger-shaped flames at various values of the initial Mach number for the axisymmetric geometry.
Figure \ref{fig-ztip.axisym.theta14} shows the scaled flame tip position
versus time for the axisymmetric geometry  for $\Theta= 14$ and the Mach
numbers $Ma=0.001 \sim 0.01$, in both linear
and logarithmic scales. It is seen that the stage of exponential
flame acceleration is more distinctive in the axisymmetric geometry
as compared to the planar case of Fig.~\ref{fig-ztip.planar.theta14}.
Figure \ref{fig-utip.axisym.theta14} is the axisymmetric analogue of Fig. \ref{fig-utip.planar.theta14}.
Similar to the planar case, the maximum flame tip
velocity strongly depends on the initial Mach number, but the
time  $\tau_{wall}$, when the maximum flame tip velocity is achieved, only depends on $Ma$  slightly.
For $\Theta = 14$ used in the simulation of Fig. \ref{fig-utip.axisym.theta14}, we have $\alpha \approx 13.5$,
and Eq. (\ref{eq63}) yields $\tau_{wall} \approx 0.15$.
The numerical simulation of Fig. \ref{fig-utip.axisym.theta14} shows somewhat delayed
maximums of flame tip velocity with $\tau_{wall} \approx 0.2$, as compared to the theoretical prediction.
This delay is attributed to the fact that for the axisymmetric case the effect of pure curvature is more pronounced
in the initial stage of finger flame propagation, since
the axisymmetric counterpart of Eq. (\ref{eq.sl.init}) yields
\begin{equation}
\label{eq.sl.init.axi}
\frac{S_{L,\,init}}{S_{L}}=1-\frac{2\,L_f}{R_{init}}=\frac{1}{2}.
\end{equation}
\noindent However, in the axisymmetric case we have a wide domain with $R=75 \, L_f$, see Sec. \ref{Sec.num.method}, which
renders pure curvature effect at the later stage of finger flame acceleration negligible for all numerical runs in  the axisymmetric case, resulting
in concurrent peaks of the flame tip velocity in Fig. \ref{fig-utip.axisym.theta14}, similar to the planar case.

\begin{figure}
\includegraphics[width=0.47\textwidth\centering]{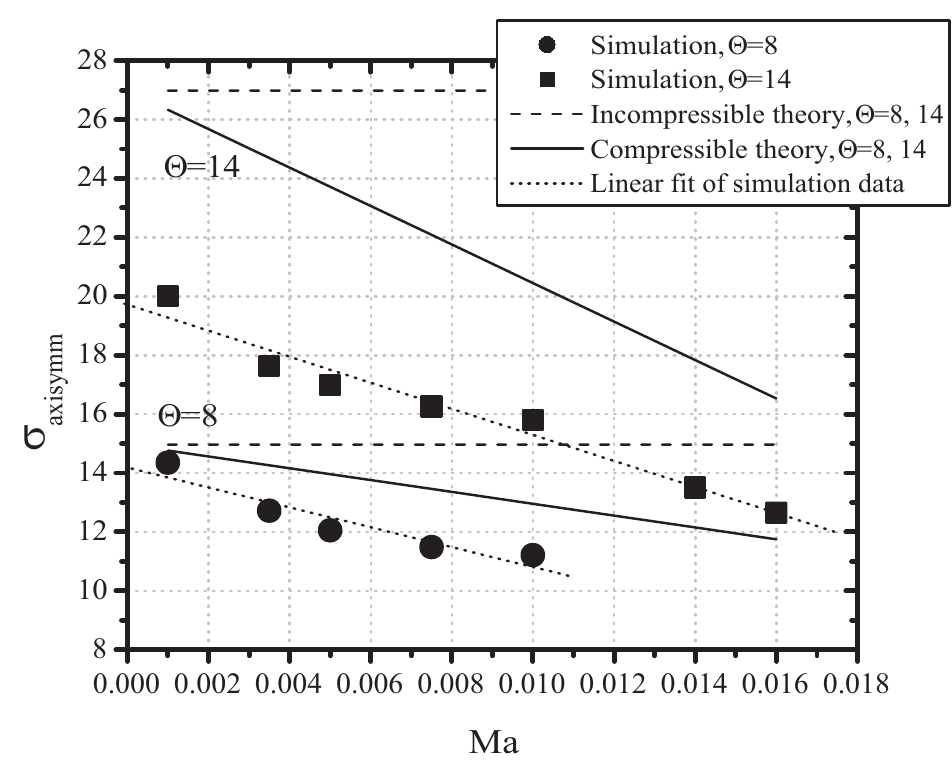}
 \caption{Scaled acceleration rate versus Mach number for axisymmetric geometry,
$\Theta=8, \
 14$. Values of $\sigma_{1}$ obtained from numerical simulations
 are shown by circles for $\Theta=8$ and by squares for
 $\Theta=14$.
 Solid lines correspond to the theoretical dependencies given by Eq.
(\ref{eq69}),
 dashed line - to incompressible theory (see Eq. (\ref{2.3.eq23-1})),
 dotted lines - to linear fit of simulation data.}
 \label{fig-sigma-Ma-axisym}
\end{figure}

 An axisymmetric counterpart of Fig. \ref{fig-utipmax.planar.theta14} is presented in Fig. \ref{fig-utipmax.axisym.theta14}.
 Contrary to the planar case, the theoretical estimate  of the maximum tip velocity shown in Fig. \ref{fig-utipmax.axisym.theta14}
is related to two opposite limiting cases descrbed in Sec. \ref{Sec.compress.axi}.
The first estimate is obtained from Eq. (\ref{eq62_III}) as
\begin{eqnarray}
& & \label{eq.utipmax.axi.begin} \frac{U_{tip,\,max}}{S_L}  =  \{2 \alpha_{1} \tanh \left( \alpha_{1} \tau_{wall} \right) - \nonumber \\
& &  - 2 Ma \, \alpha \left[1 + \gamma \left(\Theta - 1\right) \right] \}
\xi_{tip,\,wall} + \Theta_{1},
\end{eqnarray}
\noindent where the flame tip position $\xi_{tip,\,wall}$  at $\tau=\tau_{wall}$ is calculated from Eq. (\ref{eq62_IV})
while
$\sigma_{1,axi}$ and $\Theta_{1}$ are given by Eqs. (\ref{eq69}) and (\ref{eq45}), respectively.
 For the second limiting case of late stage  acceleration, the flame tip velocity estimate is
given by Eq.~(\ref{eq67}) as
\begin{eqnarray}
\label{eq.utipmax.axi.end}
\frac{U_{tip,\,max}}{S_L} = - 2 Ma ( \Theta & - & 1 ) (2\Theta \gamma  -  \gamma + 1) \xi_{tip,\,wall}^{2} + \nonumber \\
& + & \sigma_{1,axi} \xi _{tip,\,wall} + \Theta _{1}, 
\end{eqnarray}
\noindent where the flame tip position $\xi_{tip,\,wall}$  at $\tau=\tau_{wall}$ is calculated from Eq. (\ref{eq74}).
We see that results of the numerical simulation in most cases are located in between the values given by Eqs. (\ref{eq.utipmax.axi.begin})
and (\ref{eq.utipmax.axi.end}), as suggested in Sec. \ref{Sec.compress.axi}, although, 
 this tendency appears to change for higher values of $Ma$.

Figure \ref{fig-sigma-Ma-2D} shows the scaled acceleration rate
versus Mach number for the planar geometry, with $\Theta=8,\ 14$. It is seen
that the agreement between theory and simulations is significantly better for
$\Theta=8$, indicating that the role of gas compressibility  increases both with $Ma$ and $\Theta$.
Furthermore, Fig. \ref{fig-sigma-Ma-2D} shows that the growth rate $\sigma$ significantly decreases with increasing $Ma$,
and the isobaric theoretical model of the finger-flame acceleration, Refs. \cite{Clanet.Searby-1996, Bychkov.et.al-2007},
noticeably overestimates the acceleration rate $\sigma$ for high-$Ma$ flames.
We next note that an accurate estimate of the acceleration rate is  of crucial importance,
e.g., for the analysis of flame-generated shocks, preheating of the fuel mixture,
pre-detonation run-up distance  and the DDT onset \cite{Bychkov-Akkerman-2006,Valiev.et.al-2008}.
Figure \ref{fig-sigma-Ma-axisym} shows the scaled acceleration rate
versus the Mach number for the axisymmetric geometry, with $\Theta=8 $ and 14.
Similar to the planar case of Fig. \ref{fig-sigma-Ma-2D}, the agreement of theory and simulations is
better for $\Theta=8$  than for $\Theta=14$, which indicates a more important role of the compressibility effects for larger $\Theta$.
For $\Theta=14$ we have even more significant deviation of the numerical data from the theoretical predictions as compared to the planar case. Still,
the trend remains the same: $\sigma$ decreases quite rapidly with  increasing $Ma$.

\begin{table}[tp]
\begin{tabular}{|c|c|c|c|c|c|c|c|c|c|}
\hline \ $\phi$ \ & \ $\Theta $ \ & \ $S_L$, cm/s \ &\ $c_s $\ m/s& \ $Ma$ \ & \ $1/t$, $\rm{s}^{-1}$ \
&\ $\sigma_1 $ \
 \\
\hline
1.0 &\ \ 8.02\ \ &\ \ 41.8\ \ &\ \ 340\ \ & 0.00123 &132.6 &\ \ 15.8\ \   \ \\
    &\ \  \ \    &\ \  \ \    &\ \      \ &         &129.7 &\ \ 15.5\ \   \ \\
\hline
\end{tabular}
\caption{Experimental values of $\sigma_{1}$ obtained from Ref. \cite{Clanet.Searby-1996}, for $\Theta \approx 8$, low-Mach
propane-air mixture and axisymmetric geometry. $\phi$ is mixture equivalence ratio, $c_s$ is sound speed in unburnt gas mixture, $1/t$ is
experimentally measured tulip flame growth rate (for two experimental runs).}
\label{table-experiment-Clanet-Searby}
\end{table}

\begin{figure}
\includegraphics[width=0.47\textwidth\centering]{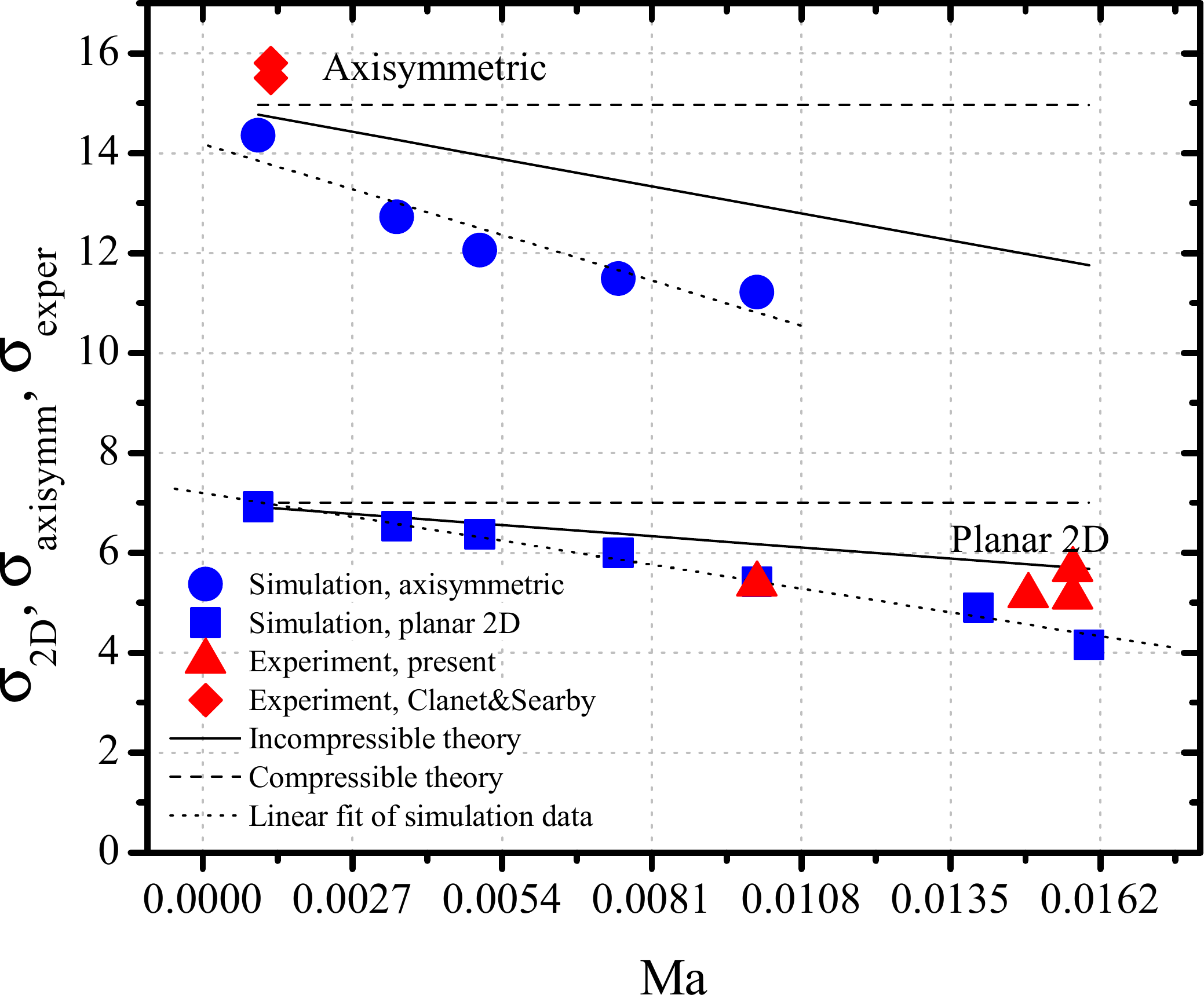}
 \caption{Comparison of scaled acceleration rates versus Mach number for
experiments and simulations for planar and axisymmetric geometries, $\Theta=8$.
 Values of $\sigma_{1}$ obtained from numerical simulations
 are shown by circles for axisymmetric geometry and by squares for
planar geometry. Triangles show experimental results of
Table~\ref{table-experiment-comparison}, diamonds - of Ref. \cite{Clanet.Searby-1996}, see Table \ref{table-experiment-Clanet-Searby}.
 Solid lines correspond to the theoretical dependencies given by Eq.
(\ref{eq45-1}) (planar) and Eq. (\ref{eq69}) (axisymmetric),
 dashed line - to incompressible theory, see Eq. (\ref{eq19-1}) (planar) and Eq.
(\ref{2.3.eq23-1}) (axisymmetric),
 dotted lines - to linear fit of simulation data.}
 \label{fig-sigma-Ma-planar-compare}
\end{figure}

Figure \ref{fig-sigma-Ma-planar-compare} shows the comparison of the
scaled acceleration rates obtained in simulations
for the planar and axisymmetric geometries, for $\Theta=8$ in both cases; the experimental results for
 $\sigma_1$ obtained in the present work and those of Ref. \cite{Clanet.Searby-1996} are also shown. The latter experimental
data is summarized
in Table \ref{table-experiment-Clanet-Searby} for two experimental runs for neutrally stable ($\phi=1.0$)
low-$Ma$, propane-air mixture with $\Theta=8.02$,
and axisymmetric geometry of the tube.
 We see from Fig. \ref{fig-sigma-Ma-planar-compare} that the scaled acceleration rate $\sigma_1$ obtained
in the present experiments for the high-$Ma$ $H_2/O_2$ stoichiometric mixture
in a channel with a quadratic cross-section are considerably closer to the planar
case  than the axisymmetric one. Thus, although the flame tip has the hemispherical shape, the global flame front acceleration is governed by the
quadratic cross section of the channel and can be described by the
 theory of Section \ref{Sec.2D.incompressible}. At the same time, values of $\sigma_1$ of Table \ref{table-experiment-Clanet-Searby}
 \cite{Clanet.Searby-1996} for low-$Ma$, propane-air mixtures are close
to the incompressible value $\sigma_0$ for the axisymmetric case, in agreement with the present theory. 
It is known that flame acceleration in
DDT sensitively depends on flame confinement and the channel geometry \cite{Wu-2012}. The present
numerical simulation and experimental results indicate that the geometry of the tube affects
the growth rate of finger flame acceleration significantly, with the axisymmetric tube 
being favorable  for faster flame acceleration due to the finger flame mechanism.

We conclude the present extensive investigation by noting that pressure jump at the flame front could potentially induce a gas velocity of $O(S_L)$,
and as such the pressure field could play an essential role in the present phenomena of interest \cite{Law-2006, Higuera-2009}. 
Here we recognize that the present theoretical results are obtained
by considering only two (quasi-planar) parts of the flame: 1) flame skirt near the closed tube end, and 2) flame tip in the vicinity of channel axis.
Due to the exponential acceleration of the flame tip, the entire process of finger flame propagation occurs rapidly. For example,
for $\Theta=8$ the characteristic non-dimensional time of reaching the maximum flame tip velocity in planar geometry is $\tau_{wall} \sim 0.25$. 
By that instant the flame tip has shifted to the non-dimensional position $\xi_{tip}=\Theta$. If we assume that the pressure field induces flame velocity of the order 
of $S_L$ at the curved parts of the flame 
(i.e. not at the skirt and the flame tip), then at the end of the finger 
flame evolution the additional displacement of the curved elements of the flame would be of the order of $S_L t_{wall} = S_L \tau_{wall} R /S_L \sim 0.25$ 
(in nondimensional units), 
i.e. considerably smaller than the flame tip displacement $\xi_{tip}=\Theta$. 
Due to the elongated flame shape
resulting from the exponential flame tip acceleration, the motion of the curved parts of the flame front does not considerably influence 
the flame acceleration. Consequently,
pressure field variation has only minimal influence on the present theory. 
It is also noted that since our problem is evolutionary (non-steady), 
the maximum flame tip velocity could be more than two orders of magnitude higher than $S_L$, hence implying that the low-Mach 
limit is not
applicable, and the pressure field cannot be considered steady as, for example, in Bunsen flame problem \cite{Higuera-2009}. 
This is beyond the scope of the present investigation.

\begin{table*}[htp!]
\begin{tabular}
{|c|c|c|c|c|c|c|} \hline \ $\Delta z_{f} / L_{f}$ \ & \  $U_{max}
/ S_L$ \ & \ $\Delta U_{max} / S_L$ \ & \ $t_{max} S_L / R$ \ & \
$\Delta t_{max} S_L / R$ \ & \ $Z_{tip, *} / R$ \ & \ $\Delta
Z_{tip, *} / R$ \
 \\
\hline
1.0 & 31.14 &  & 0.3265 &  & 4.5095 &  \\

0.5 & 34.68 & 3.54 &  0.3032 & 0.2706 & 5.282 & 0.7725 \\

0.25 & 36.171 & 1.43 &  0.2973 &  0.0059 & 5.441 & 0.159 \\

0.125 & 36.85 &  0.679  &  0.2965 &  0.0008 & 5.474 &  0.033  \\

\hline
\end{tabular}
\caption{Resolution tests for planar geometry, $Ma=0.005$,
$\Theta=8$.} \label{table-res-test}
\end{table*}

Finally, it is also noted that the influence of gas compressibility on the flame acceleration obtained in the present paper is 
qualitatively different from that of the DL instability in compressible gases and plasmas 
\cite{Travnikov-1997, Travnikov-1999, Modestov-2009}. The compressibility effect renders the DL instability 
much stronger in both linear and nonlinear stages, while in the case of finger flame acceleration 
compressibility moderates flame acceleration in tubes considerably.


\section{Conclusions}
\label{Sec.conclusions}

The theory, experiments and numerical simulations of the present work show
that the growth rate $\sigma$ of the finger-flame acceleration from the closed end of a channel/tube
decreases significantly with increasing initial Mach number, $Ma$. Hence, previous theoretical
estimates of \cite{Clanet.Searby-1996, Bychkov.et.al-2007}, derived with the incompressible approximation,
overestimate $\sigma$ for flames with high
laminar burning velocities, such as the $H_2/O_2$ or acetylene/air flames. In the present study, we account for  gas compression 
through  expansion for small Mach number up to first-order terms, and validate the theoretical analysis  by numerical simulations and experiment.
The present numerical simulation and theory show that the maximum flame tip velocity significantly depends
on $Ma$, with the scaled time  of the
maximum flame tip velocity being almost independent of it.
The present results collectively demonstrate that the geometry of the channel affects
the growth rate of the finger flame significantly, with the axisymmetric channel being more conducive for fast initial flame acceleration from the
closed end in channels with smooth walls.
It is emphasized that compressibility effects should be taken into account when estimating 
the strength of shock waves generated by the
initial finger-type flame acceleration,
 pre-heating of the unburnt fuel mixture ahead of the flame,
the DDT onset time and position.


\section{Acknowledgements}

The authors are grateful to Fujia Wu and Hemanth Kolla for useful discussions.
This work was supported by the Swedish Research Council (VR) and Stiftelsen Lars Hiertas Minne grant FO2010-1015. Numerical
simulations were performed at the High Performance Computer Center
North (HPC2N), Ume\aa, Sweden, through the SNAC project 001-10-159. 
Participation of Princeton University was supported by the US Air Force Office of Scientific Research.


\section{Appendix: Resolution and laminar velocity tests}
\label{Sec.appendix}

\begin{figure}
\includegraphics[width=0.42\textwidth\centering]{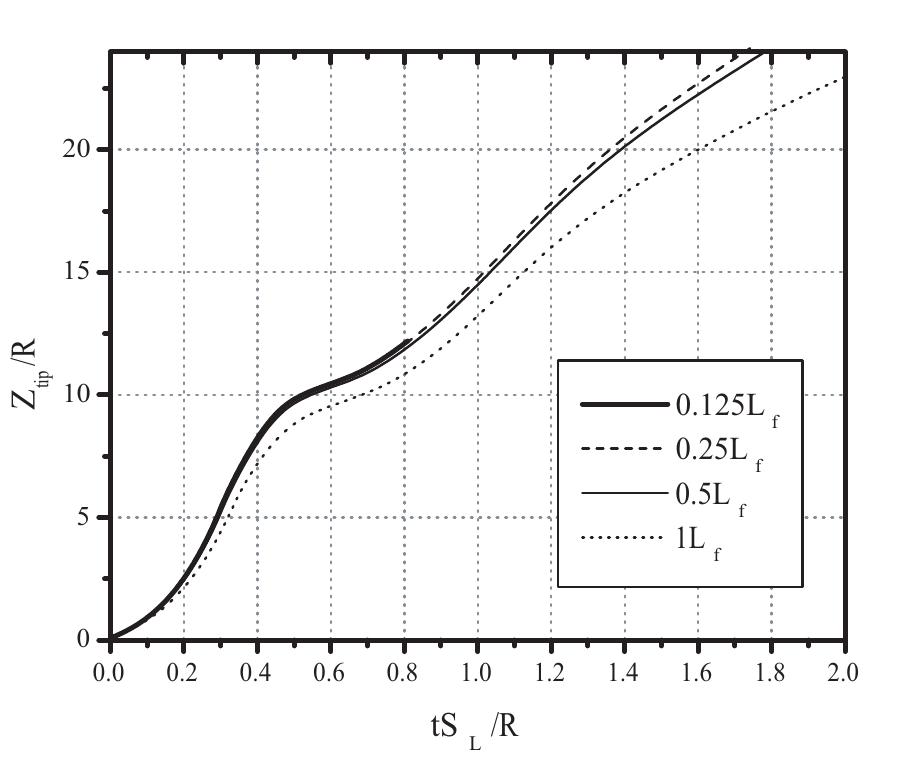}
 \caption{Flame tip position versus time for different values of
 the mesh size. Thick solid line correspond to $\Delta z_{f} / L_{f}=0.125$,
dashed line - to
 $\Delta z_{f} / L_{f}=0.25$, thin solid line - to $\Delta z_{f} / L_{f}=0.5$,
dotted line - to $\Delta z_{f} / L_{f}=1.0$}
 \label{fig-1-test}
\end{figure}

\begin{figure}
\includegraphics[width=0.42\textwidth\centering]{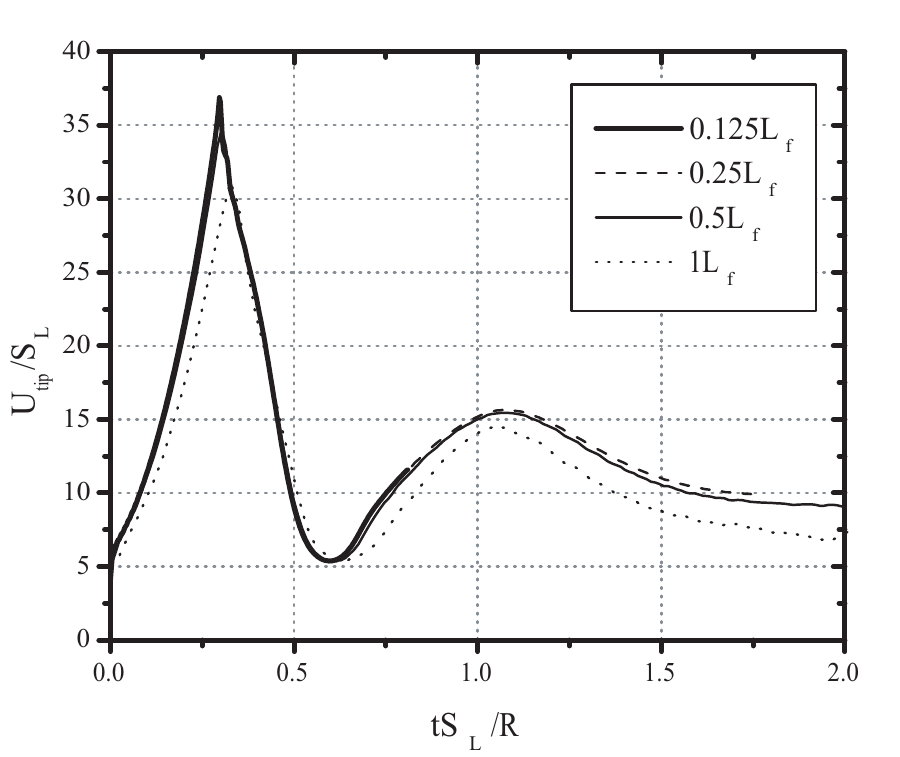}
 \caption{Velocities of the flame tip versus time for different values of
 the mesh size. Thick solid line correspond to $\Delta z_{f} / L_{f}=0.125$,
dashed line - to
 $\Delta z_{f} / L_{f}=0.25$, thin solid line - to $\Delta z_{f} / L_{f}=0.5$,
dotted line - to $\Delta z_{f} / L_{f}=1.0$}
 \label{fig-2-test}
\end{figure}

In order to check if the adopted resolution is sufficient to study
the flame acceleration process, we performed the resolution tests
for the primary results for $Ma=0.005$. The grid
size in the flame domain varied between $0.125L_{f} $, $0.25L_{f} $,
$0.5L_{f} $ and $1L_{f} $. We checked the velocity of the flame
tip at the instants corresponding to the state of the maximum flame tip
velocity, as well as the flame tip position at the time instant $t S_L/R
=0.303$. The resolution test results are presented in Table
\ref{table-res-test} and in Figs. \ref{fig-1-test} and
\ref{fig-2-test}.

Notation: $\Delta z_f / L_f $ is the spatial step in the flame
grid domain; $U_{max} / S_L $   maximum flame tip velocity
(see Fig. (\ref{fig-2-test})); $t_{max} S_L / R$   scaled
time moment corresponding to the maximum of flame tip velocity;
$Z_{tip, *} / R$   flame tip position at time $t_{*}
S_L/R =0.303$ (see Fig. (\ref{fig-2-test})). $\Delta U_{max} / S_L
$   increment of $U_{max} / S_L $ calculated in the table row
$i$ as $\Delta U_{max} (i)=U_{max} (i)-U_{max} (i-1)$. Increments
for $t_{max} S_L / R$ and $Z_{tip, *} / R$ are calculated in a
similar manner. Resolution in the wave grid domain is equal to
$\Delta z_{w} = 2\times \Delta z_{f} $ for each run.

Table \ref{table-res-test} and Figs. \ref{fig-1-test} and
\ref{fig-2-test} show good convergence of the numerical solution
with the increase of mesh resolution. Resolution tests also showed
convergence of time corresponding to the maximum flame tip velocity
with increasing resolution.

In addition to the resolution tests, laminar flame velocity tests were
performed for $Ma=0.001,\ \Theta=8,\ 14 $. The numerical setup was
similar to the main numerical experiments, except for planar
initial flame front and tube width $2R=4L_f$. That value was chosen
to be below the critical diameter needed for the growth of the
hydrodynamic instability
\cite{Liberman.et.al-2003}, so that the flame front remains
planar during the test simulation. For $\Theta=14$, the measured
laminar planar flame velocity in laboratory frame was $U_{f,\
lab}/S_L=13.97\pm 0.03$; for $\Theta=8$ $U_{f,\ lab}/S_L=7.97\pm
0.05$.



\end{document}